%% file: main.tex
\documentclass[fleqn,usenatbib]{mnras}

\usepackage{newtxtext,newtxmath}

\usepackage[T1]{fontenc}

\DeclareRobustCommand{\VAN}[3]{#2}
\let\VANthebibliography\thebibliography
\def\thebibliography{\DeclareRobustCommand{\VAN}[3]{##3}\VANthebibliography}

\usepackage{graphicx}	
\usepackage{amsmath}	
\usepackage{gensymb}
\usepackage{hhline}     
\usepackage{subcaption} 
\usepackage{circuitikz}
\usepackage{pgfplots}
\usepackage{tikz}
\usetikzlibrary{arrows.meta, positioning}
\usepackage{xcolor}

\title[Flyby kinematics]{Kinematic Signatures of Stellar Flybys in Protoplanetary Discs}

\author[V. R. Prasad et al.]{
Vasundhara R. Prasad,$^{1}$\thanks{E-mail: vprasad@ast.cam.ac.uk}
Cristiano Longarini,$^{1}$
Carolin N. Kimmig,$^{2}$
and Cathie J. Clarke$^{1}$
\\
$^{1}$Institute of Astronomy, University of Cambridge, Madingley Road, Cambridge CB3 0HA, United Kingdom \\
$^{2}$Dipartimento di Fisica, Università degli Studi di Milano, via Celoria 16, 20133, Milano, Italy
}

\date{Accepted XXX. Received YYY; in original form ZZZ}

\pubyear{\the\year{}}

\begin{document}
\label{firstpage}
\pagerange{\pageref{firstpage}--\pageref{lastpage}}
\maketitle

\begin{abstract}
Recent high-resolution molecular line observations of protoplanetary discs have revealed large-scale non-Keplerian features in the gas kinematics of several systems, which can be interpreted as signatures of warps in the disc. Warps can arise from a range of physical mechanisms, and it remains unclear whether the observed kinematic residuals can be uniquely attributed to warping, or to different competing effects. Stellar flybys are one such mechanism capable of both warping a disc and directly perturbing its kinematics. Using 3D smoothed particle hydrodynamics simulations, we study the evolution of protoplanetary discs perturbed by inclined and coplanar stellar flybys, focusing on timescales of several thousand years after periastron rather than the immediate aftermath of the encounter. We find that inclined stellar flybys initially induce warping and twisting in the disc, but that this geometric perturbation is damped on a timescale set by the disc viscosity. Despite this geometric damping, synthetic ALMA observations of the CO kinematics show that complex non-Keplerian spiral features persist in the velocity residuals long after the warped morphology has relaxed, including in coplanar encounters where no warp is ever induced. Fitting these residuals with a simple warped-disc model, the degree of warping in the disc is systematically overestimated. We therefore conclude that kinematic residuals alone are insufficient to establish that a disc is warped, as flybys induce kinematic complexity independently of warps. We further show that scattered light observations provide an independent diagnostic, showing spiral arms and lateral asymmetries.
\end{abstract}

\begin{keywords}
protoplanetary discs -- planets and satellites: formation -- hydrodynamics -- methods: numerical.
\end{keywords}



\section{Introduction}

The study of protoplanetary discs has in recent years been revolutionised by observations made using the Atacama Large Millimeter/sub-millimeter Array (ALMA). The unprecedented spatial and spectral resolution of ALMA has enabled us to study the dust and gas emission from protoplanetary discs. Notably, high-resolution observations of molecular lines in the gas disc has enabled us to probe the structure and kinematics of the disc \citep{Pinte_2023}. 

The ALMA large program exoALMA \citep{exoALMA} has leveraged these capabilities to study fifteen sources' kinematic structures in order to identify young planets. Analysis of the \textsuperscript{12}CO and \textsuperscript{13}CO $J=3-2$ emission lines has revealed large-scale deviations from Keplerian rotation in the disc kinematics \citep{Izquierdo_2025,Stadler_2025,Zawadzki_2025}. Many of these large-scale features are azimuthal asymmetries resembling $m=1$ spirals in the line-of-sight (LOS) velocity fields, most notably in the discs MWC 758 and CQ Tau.  Following the work of \citet{Young_Alexander_Rosotti_Pinte_2022}, who find that small warps in the disc can produce similar features, \citet{Winter_2025} suggest that the observed large-scale velocity features  can be interpreted as warps in the disc.

The geometry of protoplanetary discs is not always flat: their orientation might change as a function of radius, and different annuli can be misaligned with respect to each other. Such discs are called warped. Warps modify the transport of angular momentum across the disc and can imprint a lasting misalignment between the orbital planes of forming planets. In order to study warp profiles in discs we work mainly with two quantities: the tilt angle $\beta(R)$ and the twist angle $\gamma(R)$, defined as 

\begin{align}
    \beta(R) &= \arccos l_z(R), \label{eqn:beta_def} \\
    \gamma(R) &= \arctan \left(\frac{l_y(R)}{l_x(R)}\right)
\end{align}

\noindent where $l_x, l_y$ and $l_z$ are the components of the unit angular momentum vector $\mathbfit{l}$ in a set of Cartesian coordinates aligned with the unperturbed plane of the disc. The tilt angle $\beta(R)$ defines the angle between $\mathbfit{l}$ and the Cartesian $z$-axis; we consider a disc warped if $\beta(R)$ varies with (spherical) radius $R$. A warped disc is additionally also considered twisted if the twist angle about the $z$-axis $\gamma(R)$ varies with radius. Note that the twist angle $\gamma$ is not defined when $\beta \approx 0$.

We can also define the warp amplitude $\psi(R)$,

\begin{equation}
    \psi(R) = \left|\frac{\mathrm{d}\mathbfit{l}}{\mathrm{d}\ln R}\right|,
\end{equation}

\noindent which is often used to characterise the degree of warping in the disc.

We can then convert these physical disc-frame quantities into small perturbations in the inclination $i$ and position angle PA of the disc when projected onto the plane of the sky,

\begin{align}
    \beta(R) &= \sqrt{\delta i(R)^2 + \delta\mathrm{PA}(R)^2\sin^2(i_0)}\\
    \gamma(R) &= \arctan\left(\frac{\delta i(R)}{-\sin(i_0)\delta\mathrm{PA}(R)}\right)
\end{align}

\noindent where $i_0$ is the fixed unperturbed inclination of the disc.

\citet{Winter_2025} use a simple analytical prescription for the kinematic features caused by warping in the disc. Using an approach that assumes all non-Keplerian features in the disc can be attributed to warps, they model the residual in the LOS velocity field, $\delta v_\mathrm{los}$ when smooth Keplerian rotation has been subtracted off, as 

\begin{equation}
    \delta v_\mathrm{los}(R,\phi) = \delta i(R)v_\phi(R)\cos i_0\cos\phi + \delta\mathrm{PA}(R)v_\phi(R)\sin i_0\sin\phi
    \label{eqn:vlos_warp}
\end{equation}

\noindent where $v_\phi(R)$ is the Keplerian azimuthal velocity. 

Physically, warps in protoplanetary discs may be caused by many different scenarios: these include, but are not limited to, the late infall of material onto the disc \citep{Dullemond_2019,Kuffmeier_2021,Krieger_2024}; a bound companion object (either planetary or stellar) on an inclined orbit with respect to the disc \citep{Facchini_2014,Young_2023}; or an encounter with an unbound stellar companion \citep{Xiang-Gruess_2016,
Cuello_2019,Cuello_2023,Kimmig_2026}. Here we focus on stellar flybys.

Stellar flybys are commonly defined as encounters with unbound perturbing stars on parabolic or hyperbolic trajectories where the distance of closest approach is less than 1000 AU \citep{Cuello_2023}. \citet{Clarke_Pringle_1993} found that flybys can lead to tidal stripping, capture of disc material by the perturber, disc truncation and the formation of spiral substructures within the disc; they also found that in the case that the perturbing star's orbit is not coplanar with the initial plane of the disc (i.e. an inclined flyby), the flyby induces a warp in the disc. \citet{Xiang-Gruess_2016} performed smoothed particle hydrodynamics (SPH) simulations of discs perturbed by single flyby encounters on inclined orbits. They determined that retrograde inclined flybys could strongly re-orient the disc by up to $60\degree$. In a detailed theoretical study, \citet{Cuello_2019} conducted SPH simulations of protoplanetary discs perturbed by stellar flybys with a wide range of parameters, and found significant warping of the disc in inclined prograde and retrograde encounters. \citet{Cuello_2020} studied the observational signatures of flyby-perturbed discs. They found that in synthetic observations of the CO kinematics, the characteristic bridge of material connecting the primary and perturbing star is readily seen in moment 0 maps. They also found spirals in the moment 1 (velocity) maps of the disc. It has previously been shown that stellar flybys can generate spiral features in the gas disc \citep{Clarke_Pringle_1993,Cuello_2019,Smallwood_2023,Prasad_2025}; it is hence very promising that these substructures can be observed in the gas kinematics. \citet{Cuello_2020} also find that spirals (and bridges in prograde encounters) can be identified in the scattered light emission from the small dust grains in the disc, as these dust grains closely follow the dust distribution in the disc.

Recently, \citet{Kimmig_2026} conducted grid-based hydrodynamical simulations of discs perturbed by inclined flyby encounters, in order to study the warp induced in the disc by the encounter. They find that the warp can persist in the disc long after the periastron of the perturbing star -- as many as 100 orbital timescales at the disc's outer edge. They also find that flybys induce spiral arms in the gas disc, in agreement with earlier studies, but that the flyby-induced warp is much longer-lived than the spirals. They suggest that both warps and spirals may both leave imprints in the disc's kinematic structure.

Linear theory \citep{PL_1995,Lubow_Ogilvie_2000} suggests that warps damp on timescales inversely proportional to the disc's viscosity, i.e. $\tau_\mathrm{damp}=1/(\alpha\Omega_\mathrm{out})$ where $\alpha$ is the disc's viscosity parameter \citep{Shakura_Sunyaev_1973}) and $\Omega_\mathrm{out}$ is the Keplerian orbital frequency at the outer edge of the disc. \citet{Kimmig_Dullemond_2024} validated this estimate using three-dimensional grid-based simulations. They recovered a strong dependence of warp dissipation on the disc viscosity, with a viscosity $\alpha\approx 10^{-2}$ (i.e. an order of magnitude larger than typical protoplanetary disc viscosities) damping the warp ten times faster than a viscosity of $\alpha\approx 10^{-3}$. The twist angle profile of the disc is also damped much faster for higher viscosities.

With this context, this work studies the kinematic signatures of protoplanetary discs perturbed by inclined and coplanar flyby encounters. We quantify the degree of warping induced by inclined flybys. We aim to study the warp's detectability in the disc's gas kinematics and to identify additional kinematic features induced by flybys. Unlike previous studies on the observational signatures of stellar flybys, we focus mainly on longer-lasting imprints of flybys on discs, rather than the signatures of flybys that are actively occurring or very recent. Noting that the high viscosities required in SPH may effect the damping of the warp induced by an inclined flyby encounter, we do not aim to make a quantitative estimate of the warp's evolution; we instead focus on the imprint of the warp on the disc's kinematics. We also examine potential observational signatures of flybys in the scattered light and dust thermal emission.

This paper is structured as follows. Section \ref{sec:methods} discusses the hydrodynamical code and simulation set-up, as well as the post-processing pipeline used to create and analyse the synthetic observations from the simulation output. Section \ref{sec:results} presents the results of the simulations and of the post-processing pipeline. In section \ref{sec:discussion} we discuss our findings and in section \ref{sec:conclusions} we draw our conclusions.

\section{Numerical Methods} \label{sec:methods}

\subsection{SPH simulations}

We model protoplanetary discs perturbed by flyby encounters using the three-dimensional SPH code \textsc{Phantom} \citep{Price_2018}. SPH (for a review see \citealt{Price_2012}) is well-suited to simulating flyby encounters because there is no preferred geometry and angular momentum is accurately conserved; however, because of the high artificial viscosity used in SPH, a degree of caution will be required in interpreting the evolution of flyby-induced warps. \textsc{Phantom} has been extensively tested for the case of parabolic flyby encounters \citep{Cuello_2019,Ménard_2020,Smallwood_2023,Smallwood_2024,Nealon_2025,Prasad_2025}. We simulated five inclined prograde and retrograde flyby encounters using \textsc{Phantom}, varying the inclination of the perturbing star's orbit relative to the circumprimary disc. A summary of the simulations is given in Table \ref{tab:simulations}.

\subsubsection{Circumprimary disc setup}

We set up a coplanar (midplane at $z=0$), initially unwarped circumprimary disc orbiting a star of mass $1 M_{\sun}$ modelled as a sink particle. The disc has inner radius $R_\text{in} = 10 \text{ AU}$ and outer radius $R_\text{out} = 150 \text{ AU}$. The disc has a total mass of $0.01 M_{\sun}$ and is flared with an aspect ratio of $H/R = 0.05$ at $R=R_\text{in}$, increasing to $H/R = 0.1$ at $R=R_\text{out}$. The accretion radius for the primary star is set at $R_\text{acc}=10 \text{ AU}$. The equation of state of the disc is chosen to be locally isothermal (following \citealt{Lodato_Pringle_2007}) and has a sound speed 

\begin{equation}
    c_s \propto R^{-q},
\end{equation}

\noindent where $q=0.25$. The value of the sound speed is set by a locally isothermal disc temperature profile given by 

\begin{equation}
    T (R) = 64.0 \text{K} \left (\frac{R}{R_\text{in}} \right)^{-0.5}.
\end{equation}

The surface density profile of the disc is given by \citep{Lodato_Pringle_2007}

\begin{equation}
    \Sigma (R) = \Sigma_0 \left (\frac{R}{R_\text{in}} \right)^{-p} \left( 1 - \sqrt{\frac{R_\text{in}}{R}}\right),
    \label{eqn:surface_density}
\end{equation}

\noindent where $p=1$ and $\Sigma_0 = 17.1 \text{ g/cm}^2$ is a normalisation constant set by the disc's mass. The effects of disc self-gravity are ignored throughout this work. We adopt a mean \citet{Shakura_Sunyaev_1973} viscosity parameter $\alpha_\text{SS} = 0.005$ by setting a fixed ``shock capturing viscosity'' $\alpha_\text{AV} = 0.26$ and following the prescription in \citet{Lodato_Price_2010},

\begin{equation}
    \alpha_\text{SS} \approx \frac{\alpha_\text{AV}}{10}\frac{\langle h \rangle}{H}
\end{equation}

\noindent where $\langle h \rangle$ is the mean smoothing length.

We set up the disc with $10^6$ Lagrangian SPH particles. We use a single set of SPH particles to represent a mixture of both gas and dust in the ``one fluid'' (or ``dust-as-fluid'') algorithm for computing dust dynamics \citep{Price_Laibe_2015}.  In the Epstein drag regime the Stokes number, which describes the aerodynamical coupling of gas to dust, is given by \citep{epstein1924}

\begin{equation}
    \text{St} = t_s \Omega_k = \frac{\rho_\text{gr}s}{\rho c_s}\Omega_k,
    \label{eqn:stokes}
\end{equation}

\noindent where $t_s$ is the stopping time (the timescale on which dust grains couple to the flow of the surrounding gas), $\Omega_k$ is the Keplerian angular frequency, $\rho$ is the total local density of gas and dust combined, $s$ is the dust grain size and $c_s$ is the gas sound speed. The one-fluid algorithm simulates dust particles tightly coupled to the surrounding gas by setting up SPH particles that are each a mixture of dust and gas an devolving relative quantities. It is therefore most suitable for particles with low Stokes numbers. In tests performed by \citet{Cuello_2019}, it was found that the one-fluid method is suitable for flyby simulations with Stokes numbers smaller than $\mathrm{St} \approx 1$. (Note that St is only strictly defined for dust in a planar, unwarped disc, but the one-fluid method is still suitable in our warped disc simulations since the stopping times of the dust particles used are sufficiently small.) In these simulations, we set up a dust disc with the same initial surface density profile (following Equation \ref{eqn:surface_density}) and radial extent as the gas disc, but with a mass of $10^{-4} M_{\sun}$ (i.e. an initial dust to gas ratio of 0.01). We choose an intrinsic dust grain density $\rho_\text{gr} = 3 \text{ g/cm}^3$ and simulate five different dust grain sizes $s$ on a logarithmic scale ranging from $10^{-4}$ cm to $0.1$ cm, adopting a standard Mathis–Rumpl–Nordsieck power law \citep*{MRN_77} where the number density $n(a)$ of grains with size $a$ obeys

\begin{equation}
    n(a) \propto a^{-3.5}.
\end{equation}

\subsubsection{Perturber setup} \label{sec:perturber_setup}

We place the perturbing star on a parabolic orbit with a periastron distance $r_\text{peri}$ of 200 AU; for a disc with initial extent of 150 AU, this is classified as a grazing encounter, perturbing the disc significantly without destroying it. The perturbing star arrives from the negative $y$-direction, leaves towards the negative $y$-direction, and is closest to the primary star (periastron) on the $y$-axis (i.e. the line $x=0$). The perturbing star's initial distance from the primary star is set to be 10 $r_\text{peri}$, and the encounter is evolved until the distance between the two stars is once again 10 $r_\text{peri}$. We set the accretion radius for the perturbing star at $1 \text{ AU}$.

We vary the inclination of the orbit around the $y$-axis in steps of $45\degree$ from $0\degree$ (coplanar prograde orbit) to $180\degree$ (coplanar retrograde orbit). In all five flyby orbits, the periastron of the perturbing star is in the plane of the disc. We choose a perturber of equal mass to the primary star ($1 M_{\sun}$).

\begin{table}
\centering
\caption{Setup of the five hydrodynamical simulations.}
\label{tab:simulations}
\begin{tabular}{cc}
    \hline
    Simulation ID & Flyby inclination $[\degree]$ \\ \hhline{==}
    \texttt{beta0} & 0\\ \hline
    \texttt{beta45} & 45 \\ \hline
    \texttt{beta90} & 90 \\ \hline
    \texttt{beta135} & 135 \\ \hline
    \texttt{beta180} & 180 \\ \hline
\end{tabular}
\end{table}

\subsection{Radiative transfer modelling} \label{sec:radiative_transfer}

We post-process select \textsc{Phantom} output files with the Monte Carlo radiative transfer code \textsc{mcfost}  \citep{Pinte_2006,Pinte_2009}. \textsc{mcfost} takes the positions of one-fluid SPH particles from the hydrodynamical simulation as input and creates a Voronoi mesh with each cell in the mesh corresponding to an SPH particle. The code then computes the temperature structure, assuming the dust is in local thermal equilibrium with the gas, as well as the molecular line emission from the CO and \textsuperscript{13}CO $J=3-2$ transitions. We use $N_\gamma = 10^8$ photon packets in order to compute the temperature structure, assuming local thermal equilibrium. We assume the disc is at a distance $140$ pc from the observer at a position angle (PA) of $0\degree$, and that it is inclined at $30 \degree$ to the line of sight of the observer. We set the CO abundance to $10^{-5}$ relative to H and the \textsuperscript{13}CO abundance to $10^{-5}/70$ relative to H (following the relative abundances of \citealt{Booth_2019}). 

The CO and \textsuperscript{13}CO cubes from \textsc{mcfost} were then processed using the \texttt{python} package \texttt{pymcfost} to produce synthetic ALMA observations. The final cubes had a spectral resolution of $100 \mathrm{ms}^{-1}$ and were convolved with a 0.1 arcsec Gaussian beam to match the exoALMA observations. The continuum emission was also removed in order to isolate the molecular line emission. 

In order to create synthetic observations of the dust continuum and scattered light emission, \textsc{mcfost} uses a ray-tracing method to produce images from the computed temperature structure of the disc. To produce synthetic ALMA dust continuum observations, we convolve the images with a 0.1 arcsec Gaussian beam and add a noise level of $5 \times 10^{-5}$ Jy/beam. To produce synthetic VLT/SPHERE scattered light observations, we convolve the image with a Gaussian beam corresponding to the 8.2m diameter of VLT'S Unit Telescope 3. To produce synthetic JWST scattered light observations, we utilise the \texttt{python} package STPSF \citep{stpsf}, which computes the full point spread function (PSF) for JWST, and then convolve the image with a kernel corresponding to this PSF. 
We produce images of the polarised intensity ($Q_\phi$) of scattered light emission. 

\subsection{Modelling channel maps and fitting warp profiles} \label{sec:discminer_warpfitter}

We employ the \textsc{discminer} channel-map modelling and analysis framework \citep{discminer} to analyse the synthetic observations generated by \texttt{pymcfost}. \textsc{discminer} assumes a smooth Keplerian model for the disc's rotation and models the inclination $i$ with respect to the observer, the position angle PA and and the offset $x_c$ and $y_c$ as free parameters. The line peak intensity, line width and line slope in the radial and vertical directions are parameterised with simple power laws in cylindrical coordinates $(R,z)$. Both the upper and lower emitting surfaces are modelled. A summary of the \textsc{discminer} attributes and their parametrisations is shown in \autoref{tab:discminer_attributes}. \textsc{discminer} is coupled with the \textsc{emcee} ensemble sampler for Markov Chain Monte Carlo (MCMC; \citealt{emcee}). We run an MCMC chain of 10,000 iterations with 100 walkers in order to obtain a set of best-fit parameters when modelling the synthetic data cube as a smooth Keplerian disc.

We can then use these best-fit parameters to produce moment maps of the peak intensity and centroid velocity. We also produce maps of the residual peak intensities and centroid velocities when comparing the synthetic observation to \textsc{discminer}'s best-fit Keplerian model. We can then fit the observed kinematic structures in the residual LOS velocity to the simple analytical warped disk model presented by \citet{Winter_2025}, using their \texttt{python} package \texttt{warpfitter} to extract a radial warp profile from the velocity residuals. 

\begin{table}
    \centering
    \caption{Attributes adopted in the \textsc{discminer} models for this work. $G$ is the gravitational constant;
    $D_0=100\,\mathrm{au}$ is a normalisation factor.
    In disc coordinates, $z$ is the height above the disc midplane,
    $R$ is the cylindrical radius, and $r$ is the spherical radius.
    The remaining variables are free parameters.
    All free parameters involved in each attribute are modelled independently, even if they are named alike.}
    \label{tab:discminer_attributes}
    
    \begin{tabular}{|l|l|}
    \hline
    \textbf{Attribute} & \textbf{Prescription} \\
    \hhline{==} 
    Orientation &
    $i,\ \mathrm{PA},\ x_c,\ y_c$ \\
    \hline
    
    Velocity &
    $v_{\rm rot}=\sqrt{{GM_\star}{r^3}R},\ v_{\rm sys}$ \\
    \hline
    
    Upper surface &
    $z_u=z_0\left({R/D_0}\right)^p
    \exp\!\left[-\left({R/R_t}\right)^q\right]$ \\
    Lower surface &
    $z_l=-z_0\left({R/D_0}\right)^p
    \exp\!\left[-\left({R/R_t}\right)^q\right]$ \\
    \hline
    
    Peak intensity &
    $I_p=I_0\left({R/D_0}\right)^p
    \left({z/D_0}\right)^q$ \\
    Line width &
    $L_w=L_{w0}\left({R/D_0}\right)^p
    \left({z/D_0}\right)^q$ \\
    Line slope &
    $L_s=L_{s0}\left({R/D_0}\right)^p$ \\
    \hline
    \end{tabular}
\end{table}

\begin{figure}
    \centering
    \begin{tikzpicture}[
        node distance=1.8cm,
        every node/.style={
            draw,
            rounded corners,
            minimum width=4cm,
            minimum height=1cm,
            align=center
        },
        arrow/.style={
            -{Latex[length=3mm]},
            thick
        }
    ]
    
    \node (step1) {\textsc{Phantom} output};
    \node (step2) [below of=step1] {Radiative transfer with \textsc{mcfost}};
    \node (step3) [below of=step2] {Generate synthetic observations with \texttt{pymcfost}};
    \node (step4) [below of=step3] {Channel map modelling with \textsc{discminer}};
    \node (step5) [below of=step4] {Fit warp profiles with \texttt{warpfitter}};
    
    \draw[arrow] (step1) -- (step2);
    \draw[arrow] (step2) -- (step3);
    \draw[arrow] (step3) -- (step4);
    \draw[arrow] (step4) -- (step5);
    
    \end{tikzpicture}
    \caption{The pipeline used to generate synthetic observations from the SPH simulation outputs, create channel maps and fit analytical warp profiles to the line-of-sight velocity residuals.}
    \label{fig:processing_pipeline}
\end{figure}
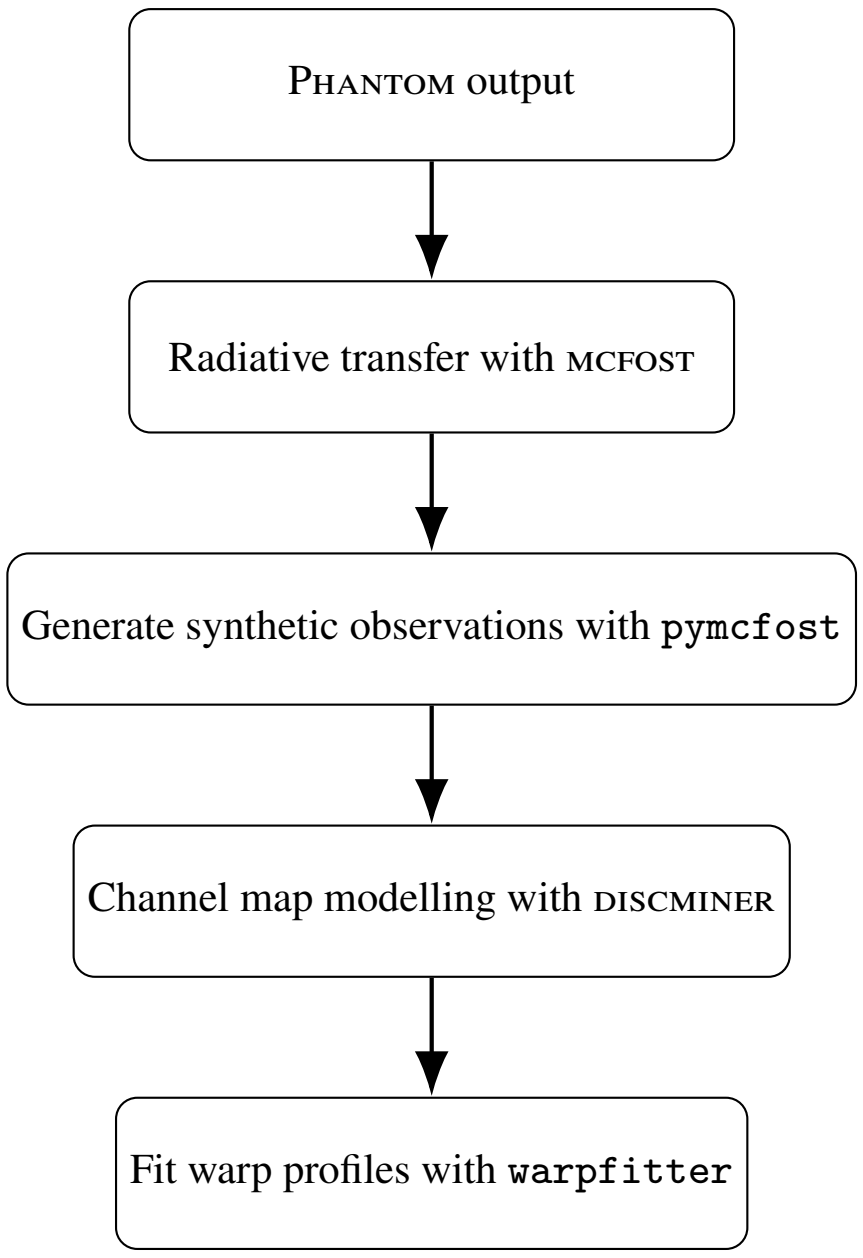

\autoref{fig:processing_pipeline} summarises the post-processing pipeline outlined in subsections \ref{sec:radiative_transfer} and \ref{sec:discminer_warpfitter}.

\subsection{Warped disc coordinates} \label{sec:warped_coords}

We expect flybys on an inclined orbit to induce global warps in the perturbed disc, such that the disc midplane is no longer the $z=0$ plane. In order to differentiate large-scale velocity perturbations in the disc (i.e. those caused by warps and changes in inclination) from additional  smaller velocity substructures (such as rings and spirals) caused by the flyby, it is necessary to construct a set of locally varying Cartesian coordinates $x', y', z'$, where the $z'=0$ plane is the midplane of the warped disc. 

In order to construct these coordinates, we must first find the unit angular momentum vector $\mathbfit{l}$ for each SPH, particle, defined as

\begin{equation}
    \mathbfit{l} = |\mathbfit{r} \times \mathbfit{v}|,
\end{equation}

\noindent where $\mathbfit{r}$ and $\mathbfit{v}$ are respectively the three-dimensional position and velocity vectors of the SPH particle relative to the central star. We then smooth out azimuthal variations in the angular momentum vector (so that the disc's warped midplane varies with radius only) by taking the azimuthal average of $\mathbfit{l}$ over concentric annuli to find the unit angular momentum vector as a function of spherical radius $R$, $\mathbfit{l}(R)$.

As the angular momentum vector points \emph{out} of the plane of the disc, and is therefore aligned with the $\hat{\mathbfit{z}}$  vector for an unperturbed disc whose midplane is in the $z=0$ plane, we can take the local value of $\mathbfit{l}(R)$ as defining the $\hat{\mathbfit{z'}}$ vector in our warped coordinate system.

In order to fully define a radially-varying, locally orthonormal system of warped coordinates, we then define

\begin{equation}
    \hat{\mathbfit{{x'}}} = |\hat{\mathbfit{x}} - \left(\hat{\mathbfit{x}}\cdot\mathbfit{l}(R)\right)\mathbfit{l}(R)|
\end{equation}

\noindent and

\begin{equation}
    \hat{\mathbfit{y'}} = \mathbfit{l}(R) \times \hat{\mathbfit{x'}}.
\end{equation}

The position of an SPH particle in the warped disc coordinates $x', y', z'$ is then trivially obtained by taking $\hat{\mathbfit{x'}}\cdot \mathbfit{r}$, $\hat{\mathbfit{y'}}\cdot \mathbfit{r}$ and $\hat{\mathbfit{z'}}\cdot \mathbfit{r}$ respectively. Using this locally-varying pseudo-Cartesian system, we can then render quantities by taking a cross-section in the midplane of the warped disc rather than in the $z=0$ plane.

\begin{figure}
    \centering
    \begin{tikzpicture}
    \begin{axis}[
        axis lines=middle,     
        enlargelimits,         
        xlabel={$x$}, ylabel={$z$},
        xtick=\empty, ytick=\empty, 
        samples=100,
        domain=-2:2,
        axis equal   
    ]
        \addplot[ultra thick, black] {tanh(x)}; 
        \addplot[thick, red, dotted] {tanh(x)};
        \addplot[very thick, blue, dotted] coordinates {(-2,0) (2,0)};
        \addplot[very thick, green, dashed] coordinates {(-2,-0.9640) (2, 0.9640)};
        \addplot[thick, black, ->] coordinates {(1,0.7616) (0.8, 1.2378)};
        \node at (axis cs:2,-0.3) [text=blue,anchor=south] {\tiny $z=0$};
        \node at (axis cs:2,1) [text=red,anchor=south] {\tiny $z'=0$};
        \node at (axis cs:-1,-0.15) [text=green, anchor=north] {\tiny $z_\mathrm{incl}=0$};
        \node at (axis cs: 0.78, 1.24) [anchor=south] {$\mathbfit{l}$};
    \end{axis}
\end{tikzpicture}
    \caption{A schematic edge-on view of a warped disc (black line) in the $x-z$ plane. As the disc is warped out of the $z=0$ plane (blue dotted line), we use the local angular momentum vector $\mathbfit{l}$ to define a pseudo-Cartesian system of coordinates and hence the midplane $z'=0$ of the warped disc (red dotted line). We can also define coordinates rotated by a constant inclination angle such that the disc's midplane $z_\mathrm{incl}=0$ (green dashed line) is not locally varying.}
    \label{fig:warp_schematic}
\end{figure}
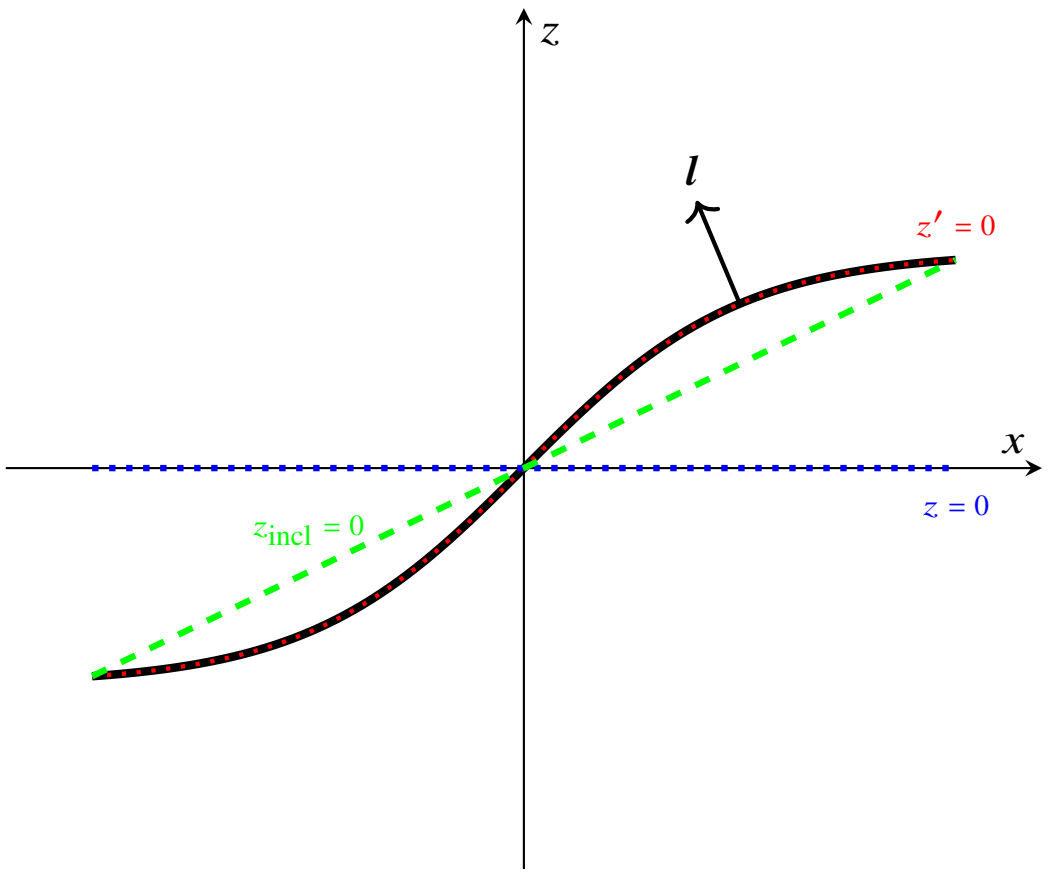

Alternatively, we can take the global value of $\mathbfit{l}$ for the \emph{entire} disc and use this to define the global plane of inclination for the disc. This is equivalent to a rigid rotation of the entire system of coordinates by a single angle, the global tilt angle of the disc. This system is most useful for viewing the velocity field when the disc is inclined but not warped. \autoref{fig:warp_schematic} illustrates the differences between the three coordinate systems.

\section{Results} \label{sec:results}

\subsection{Hydrodynamical simulations}

\begin{figure*}
    \centering
    \includegraphics{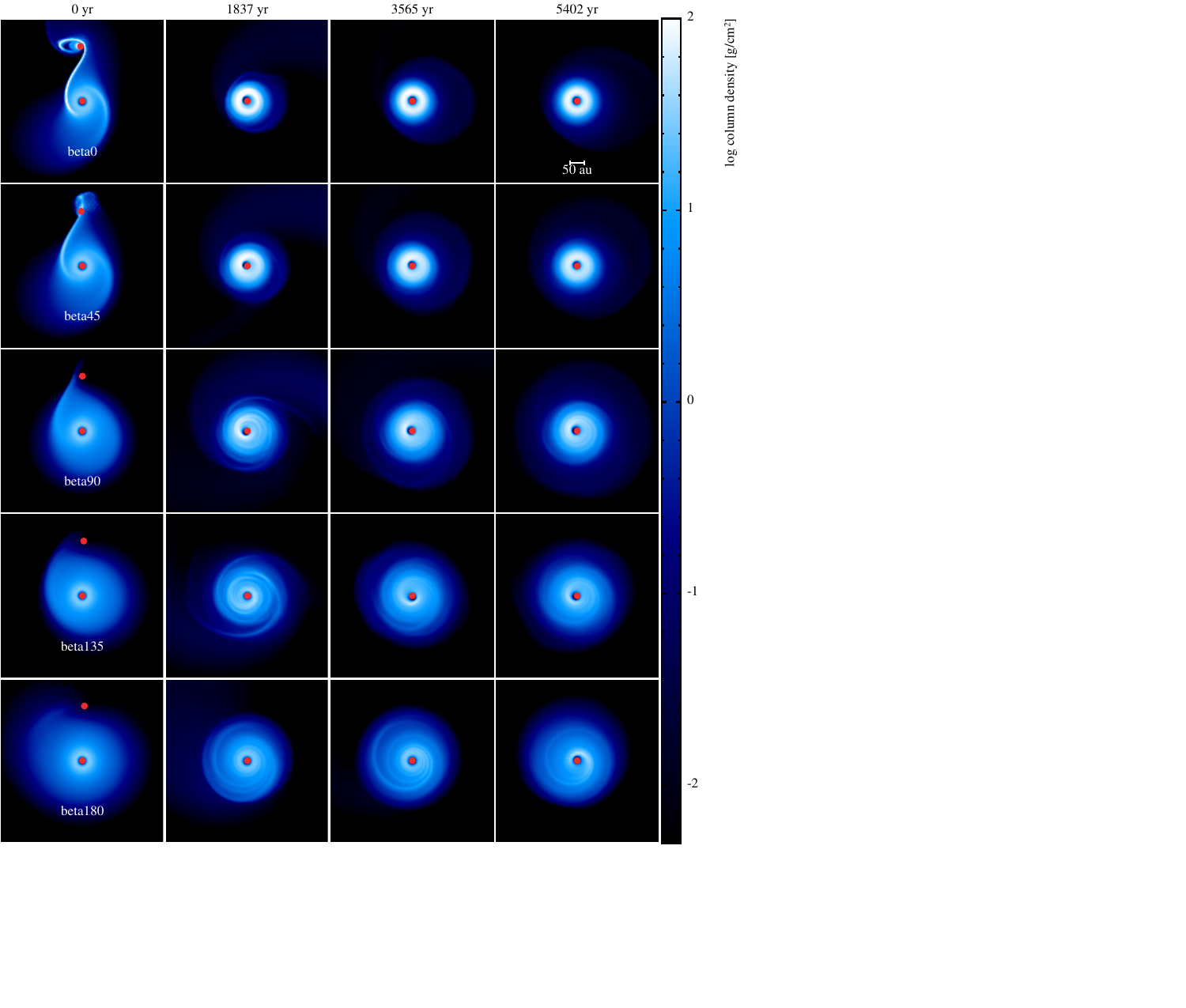}
    \caption{Face-on view (in the $x$-$y$ plane) of the gas column density of the primary disc for (top to bottom): the \texttt{beta0}, \texttt{beta45}, \texttt{beta90}, \texttt{beta135} and \texttt{beta180} simulations (as listed in \autoref{tab:simulations}). Columns (left to right) are at times 0 yr (periastron), 1837 yr, 3565 yr and 5402 yr after periastron. Sink particles (in red) are large for visualisation purposes only.}
    \label{fig:faceon_densities}
\end{figure*}

\begin{figure*}
    \centering
    \includegraphics{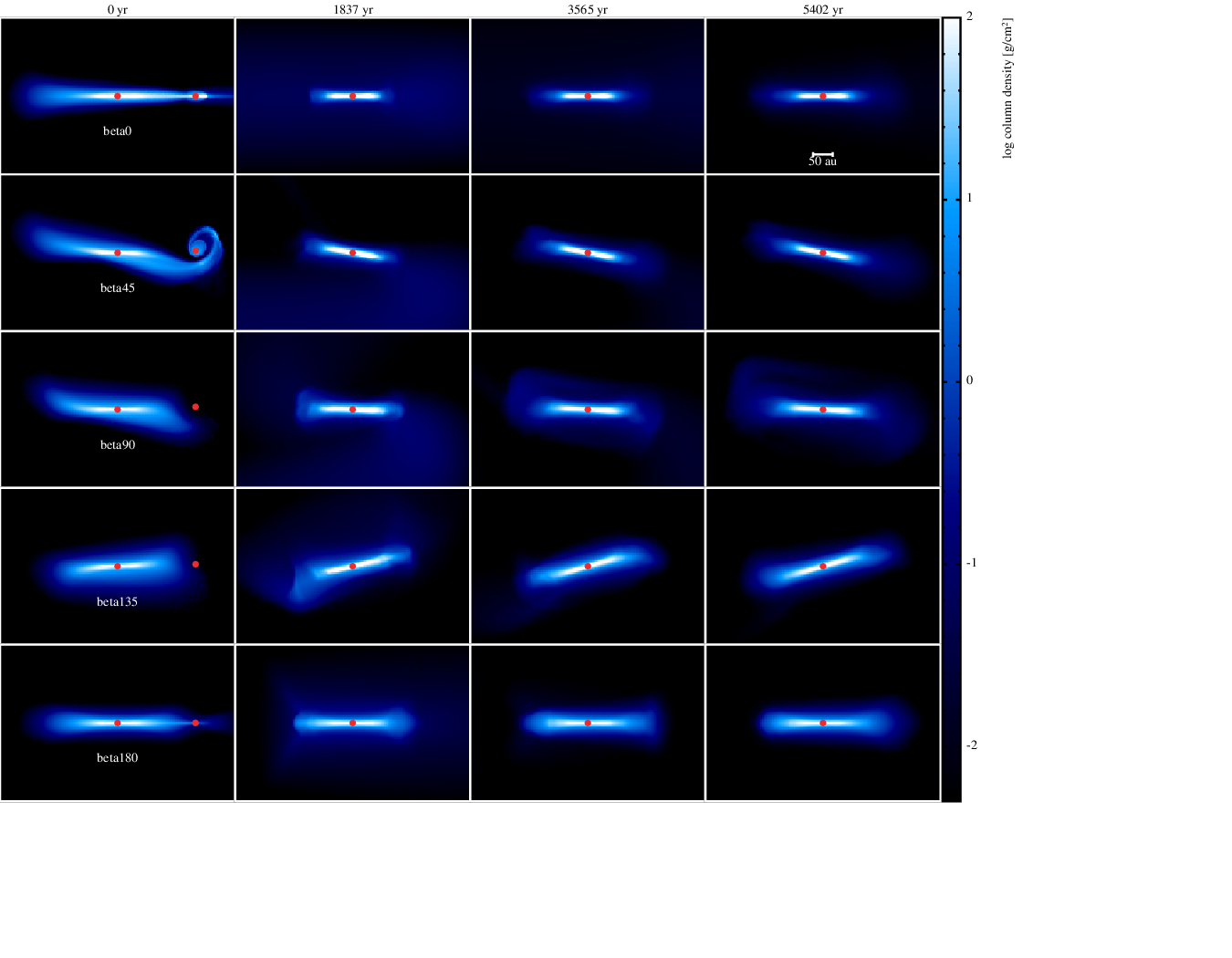}
    \caption{Edge-on view (in the $y$-$z$ plane) of the gas column density of the primary disc for (top to bottom): the \texttt{beta0}, \texttt{beta45}, \texttt{beta90}, \texttt{beta135} and \texttt{beta180} simulations (as listed in \autoref{tab:simulations}). Columns (left to right) are at times 0 yr (periastron), 1837 yr, 3565 yr and 5402 yr after periastron. Sink particles (in red) are large for visualisation purposes only.}
    \label{fig:edgeon_densities}
\end{figure*}

Figures \ref{fig:faceon_densities} and \ref{fig:edgeon_densities} show the gas column density (face-on and edge-on) in the primary disc at periastron and later times after the flyby. In \autoref{fig:faceon_densities}, we see that the gravitational perturbation of the flyby induces spirals in the gas disc. These spirals are most prominent and persist for longest in the polar and retrograde cases, \texttt{beta90}, \texttt{beta135} and \texttt{beta180}. By contrast, in the prograde flybys \texttt{beta0} and \texttt{beta45}, the gas density spirals have dissipated by 5000 years after periastron; this agrees with the findings by \citet{Prasad_2025} that gas density spirals are longer-lived in retrograde than in prograde flybys.  In \autoref{fig:edgeon_densities}, we show an edge-on view of the gas disc for each flyby. We can see that the three inclined flybys, \texttt{beta45}, \texttt{beta90} and \texttt{beta135}, leave the disc inclined with respect to its original unperturbed midplane. At periastron, the disc appears warped (i.e. its inclination varies with radius), but at later times the warp is less pronounced and the disc instead has a constant global misalignment. We can therefore see that flybys induce both large-scale warps and inclinations as well as smaller substructures like spirals.

\begin{figure*}
    \centering
    \includegraphics{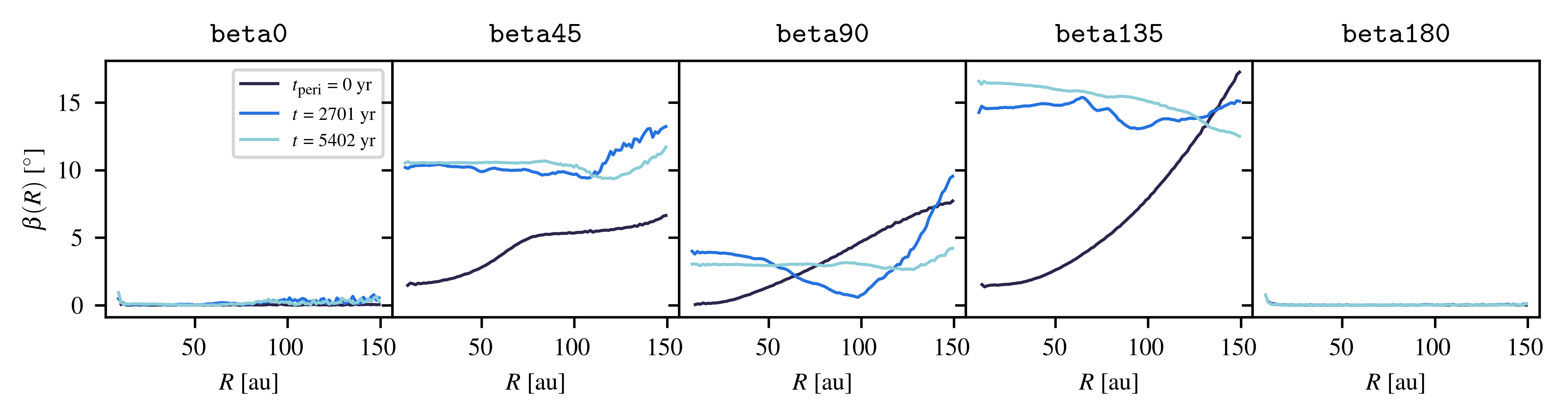}
    \caption{Evolution of the tilt angle $\beta$ as a function of spherical radius $R$ at 0 yr (periastron), 2701 yr and 5402 yr after periastron for (left to right): the \texttt{beta0}, \texttt{beta45}, \texttt{beta90}, \texttt{beta135} and \texttt{beta180} simulations.}
    \label{fig:beta_evolution}
\end{figure*}

\begin{figure*}
    \centering
    \includegraphics{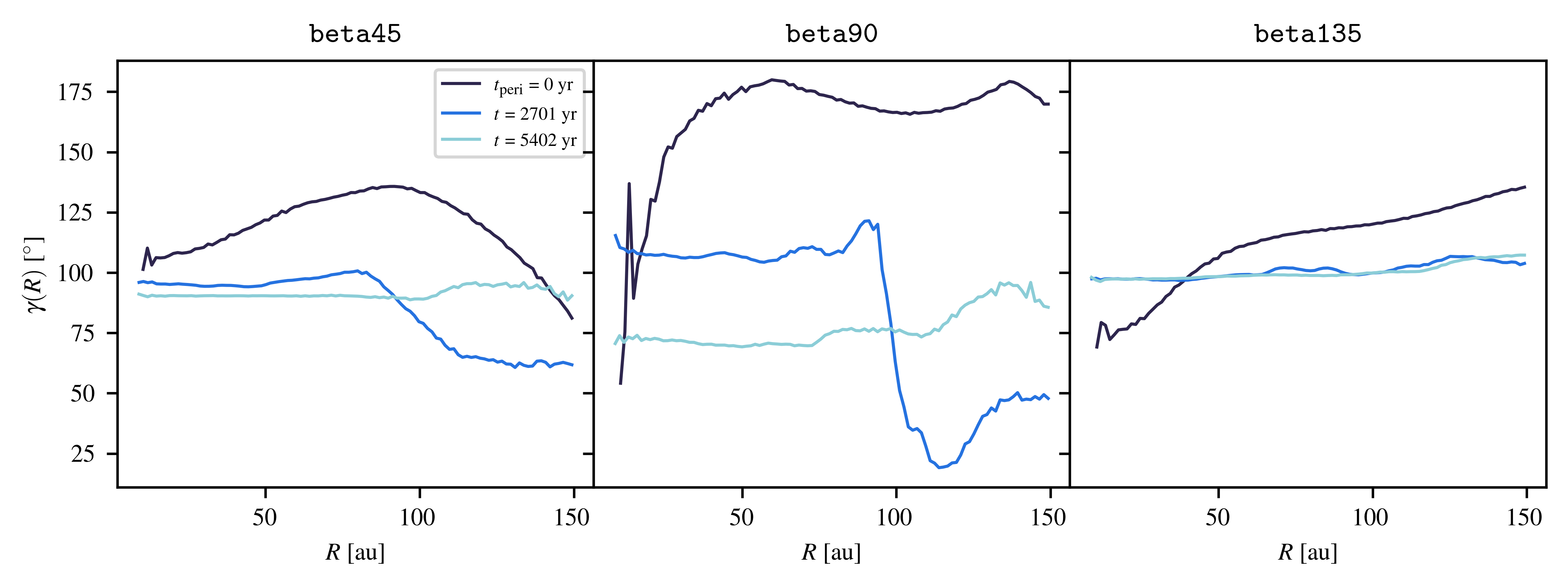}
    \caption{Evolution of the twist angle $\gamma$ as a function of spherical radius $R$ at 0 yr (periastron), 2701 yr and 5402 yr after periastron for (left to right): the \texttt{beta45}, \texttt{beta90} and \texttt{beta135} simulations.}
    \label{fig:gamma_evolution}
\end{figure*}

We can further study the evolution of the warp induced by the flyby by studying the radial profiles of the tilt angle $\beta$ and the twist angle $\gamma$ at different timesteps. \autoref{fig:beta_evolution} shows the azimuthally-averaged tilt angle $\beta(R)$ as a function of spherical radius at periastron as well as later times in the simulation for all five flyby simulations. As could be qualitatively seen in \autoref{fig:edgeon_densities}, the coplanar flybys \texttt{beta0} and \texttt{beta180} do not incline the disc, and the value of $\beta(R)$ stays constant at $0\degree$ throughout the disc. These two simulations will hence prove useful as ``control cases'' in which we can isolate the effects of spiral substructures in the kinematics from the effects of warps. By contrast, in the three inclined simulations, \texttt{beta45}, \texttt{beta90} and \texttt{beta135}, the disc is warped at periastron with the tilt angle $\beta(R)$ increasing at larger radii; this is most extreme in the inclined retrograde flyby \texttt{beta135}. At later times, the variation in $\beta(R)$ decreases, with the inner disc at constant inclination by the end of the simulation. The inclined retrograde flyby \texttt{beta135} once again induces the largest global inclination to the disc, with $\beta(R) \approx 15 \degree$ at the end of the simulation. The \texttt{beta45} and \texttt{beta90} cases have final, roughly constant inclinations of approximately $10\degree$ and $3\degree$ respectively. From the evolution of these profiles we can therefore conclude that inclined flybys initially warp the disc, but the warp has relaxed into a fixed global inclination, particularly in the inner disc, by a few thousand years after the flyby. We note that the high viscosity used in SPH simulations is likely to have caused this relatively fast relaxation of the induced warp and discuss the numerical effect of unusually high viscosities further in \autoref{sec:warp_discussion}.

\autoref{fig:gamma_evolution} shows the evolution of the azimuthally-averaged \emph{twist} angle $\gamma(R)$ as a function of spherical radius for the three inclined flybys \texttt{beta45}, \texttt{beta90} and \texttt{beta135}. (The twist angle is not well-defined for a disc that is not warped, i.e. with $\beta=0$, for which reason the $\gamma(R)$ profiles are not shown for the two coplanar flybys as they did not induce warps in the disc.) In all three inclined flyby cases, the disc is twisted at periastron with a varying value of $\gamma(R)$. However, by later times, the value of $\gamma(R)$ becomes constant (especially in the inner disc, $R<100$ AU), leaving the disc untwisted by the end of the simulations, a few thousand years after the flyby. We note again that the high viscosity used in our SPH simulations is expected to influence the timescale over which the flyby-induced twist relaxes.

\subsection{Velocity fields in warped coordinates}

\begin{figure*}
    \centering
    \includegraphics{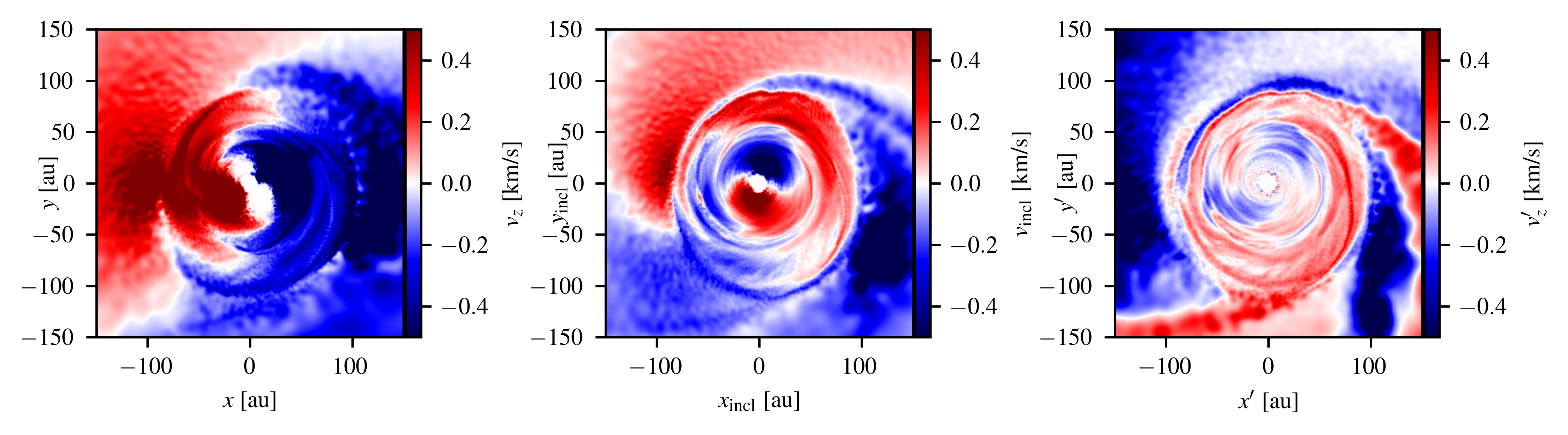}
    \caption{The vertical velocity of the disc in (left) the Cartesian $x-y$ plane, (middle) the plane of the inclined disc and (right) the locally-varying midplane of the warped disc in the \texttt{beta45} simulation at 1080 yr after periastron.}
    \label{fig:vz_comparison}
\end{figure*}

It is also illuminating to study the gas velocity of the perturbed disc. \autoref{fig:vz_comparison} shows the vertical velocities in the disc in each of the three coordinate systems described in \autoref{sec:warped_coords}, for a snapshot from the \texttt{beta45} simulation relatively soon (1080 yr) after periastron. In the left panel, with the velocity rendered in unperturbed Cartesian coordinates (i.e. with the $z=0$ plane corresponding to the plane of the unperturbed disc), we see that the vertical velocity component is dominated by the Keplerian rotation of the inclined disc, with half the disc moving towards the observer and the other half moving away from the server. In this coordinate frame it is therefore difficult to distinguish ``true'' signatures in the disc's vertical velocity from the much larger $v_\phi$ component, which is the disc's Keplerian rotation. The problem can be ameliorated by rotating these coordinates by the fixed global inclination of the disc (as in the green dashed line of \autoref{fig:warp_schematic}), which is shown in the middle panel; however, we still see a component of the Keplerian rotation in the inner disc, so rotating Cartesian coordinates by a fixed angle is not sufficient to isolate the vertical velocity component of the disc. As observed discs are generally inclined with respect to the plane of the sky, these coordinates represent the velocity field most similar to what we expect to actually observe. Finally, in the right panel the velocity is rendered in the locally-varying midplane of the warped disc (as in the red dotted line of \autoref{fig:warp_schematic}). The Keplerian rotation is now no longer visible, and instead we see spiral structures in the $v_z'$ component of the disc velocity. We therefore can see from the hydrodynamics themselves that the flyby induces substructures in the disc kinematics \emph{beyond} the warp itself, as substructures are still visible when the geometry of the warp is accounted for.

\begin{figure*}
    \centering
    \includegraphics{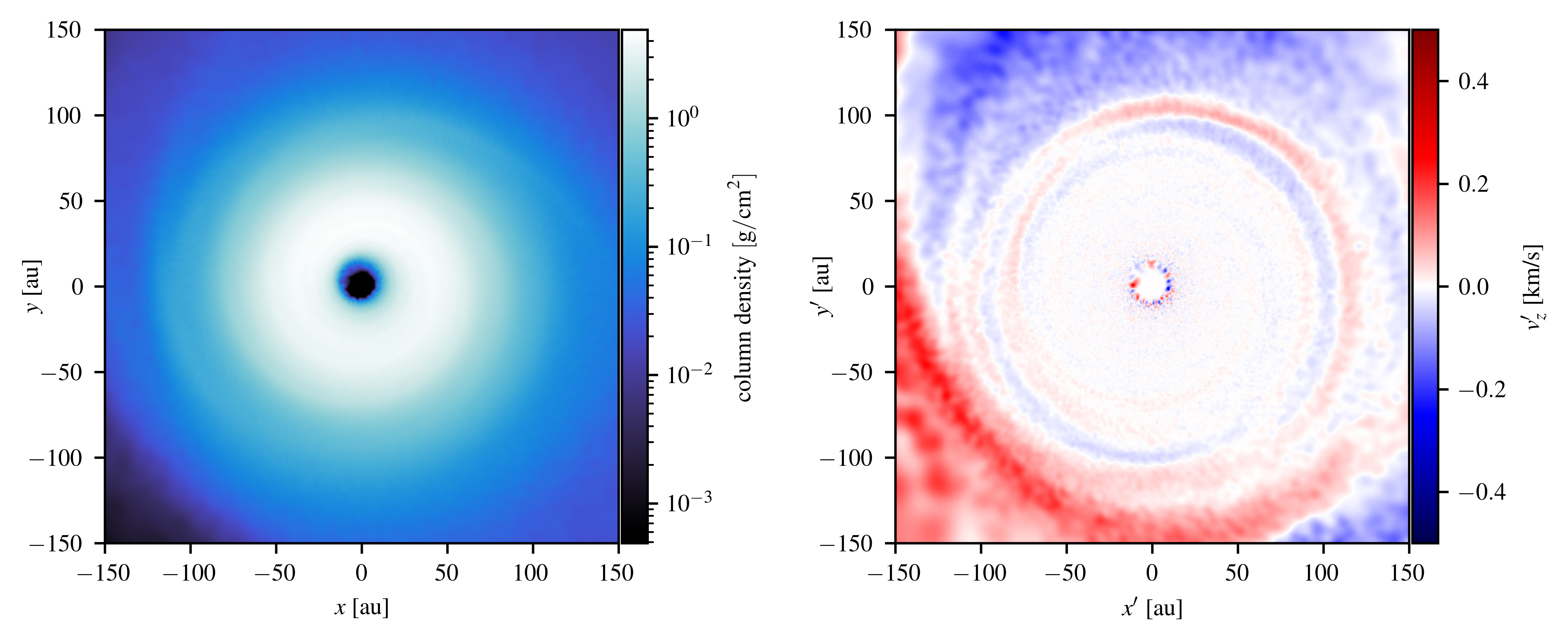}
    \caption{Snapshots of the disc 5402 yr after periastron in the \texttt{beta45} simulation. Left: column density integrated over the unperturbed Cartesian $z$-axis. Right: vertical velocity $v'_z$ in the locally-varying midplane of the warped disc.}
    \label{fig:density_vz_comparison}
\end{figure*}

In \autoref{fig:density_vz_comparison}, we show the disc density and velocity field 5402 yr after the flyby periastron. We again isolate the substructures in the velocity field caused only by the flyby by rendering them in the warped midplane of the disc, so that only the $v_z'$ component of the disc velocity perpendicular to the warped midplane can be seen. This demonstrates that kinematic substructures in the disc can still be seen long after periastron and outlast the perturbations seen in the disc's column density.

\subsection{Radiative transfer and channel map modelling}

We follow the procedure outlined in \autoref{sec:radiative_transfer} to produce synthetic ALMA observations of discs perturbed by inclined flybys. As described in \autoref{sec:discminer_warpfitter}, we then use the \textsc{discminer} channel-map modelling and analysis framework to produce maps of the centroid velocity residuals of the synthetic observations compared to a best-fit Keplerian model. \autoref{tab:discminer_table} summarises the best-fit Keplerian stellar mass, systemic velocity and orientation parameters inferred by \textsc{discminer} for both the CO and \textsuperscript{13}CO synthetic observations. In most cases, we recover the parameters expected from the hydrodynamical input and radiative transfer calculations, i.e. a stellar mass $M_* \approx 1.0 M_{\sun}$, a systemic velocity $v_\mathrm{sys} \approx 0$ km/s, and a sky-frame inclination of $i\approx 30\degree$. The only exception is the \texttt{beta135} simulation, where the MCMC best-fit value of the stellar mass is larger than expected and the sky-frame inclination is smaller. We attribute this to the fact that the \texttt{beta135} flyby encounter appears to induce the largest degree of warping and misalignment in the disc, as shown in \autoref{fig:beta_evolution}; it is therefore the most perturbed disc of all simulations and the most difficult to fit to a Keplerian model.

Throughout this and the following subsection, we focus our attention on the final snapshot of each \textsc{Phantom} simulation, 5402 yr after periastron. At this stage in the flyby encounter, the perturber is at a distance of 2000 AU (i.e. $10 r_\mathrm{peri}$) from the primary star.

\input{discminer_table_orientation}

\begin{figure*}
    \centering
    \begin{subfigure}{\textwidth}
        \includegraphics[scale=0.5]{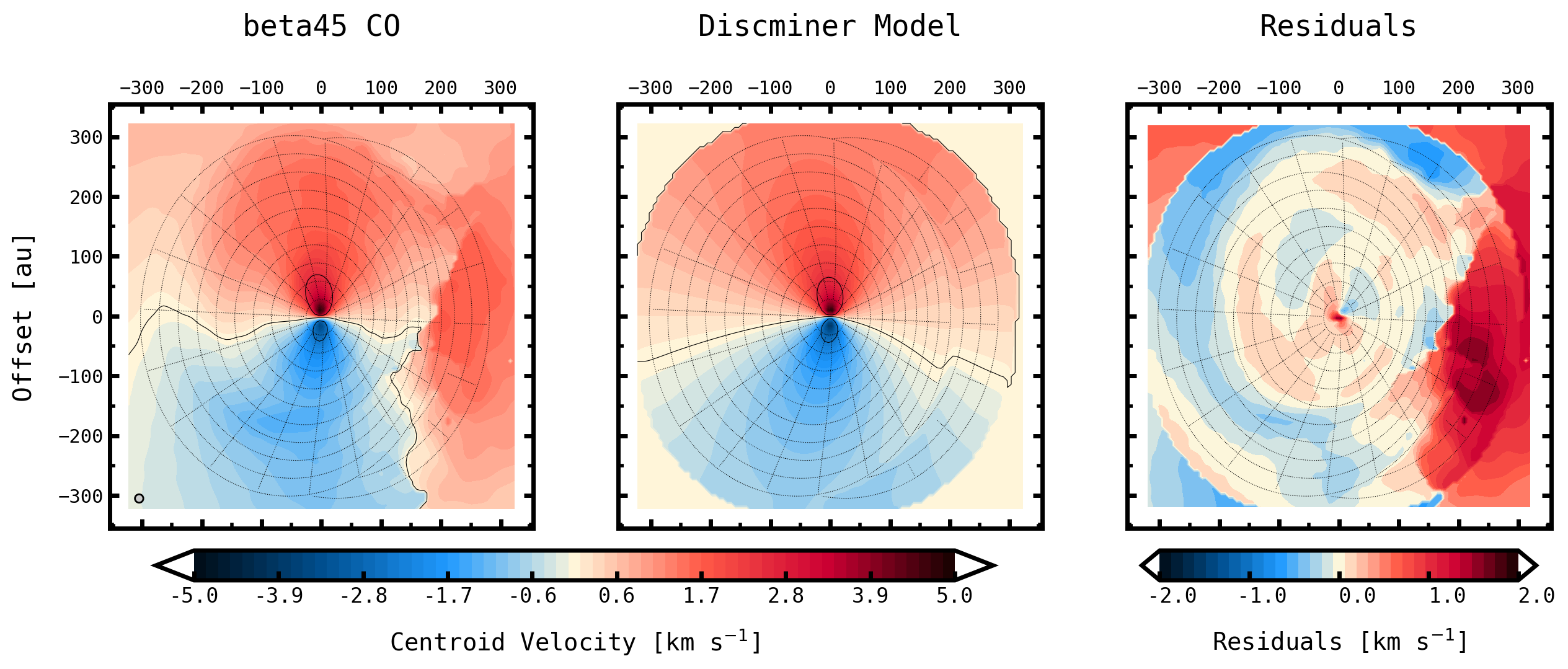}
        \caption{Velocity moment map and residuals for the CO emission.}
        \label{fig:CO_residuals}
    \end{subfigure}
    \begin{subfigure}{\textwidth}
        \includegraphics[scale=0.5]{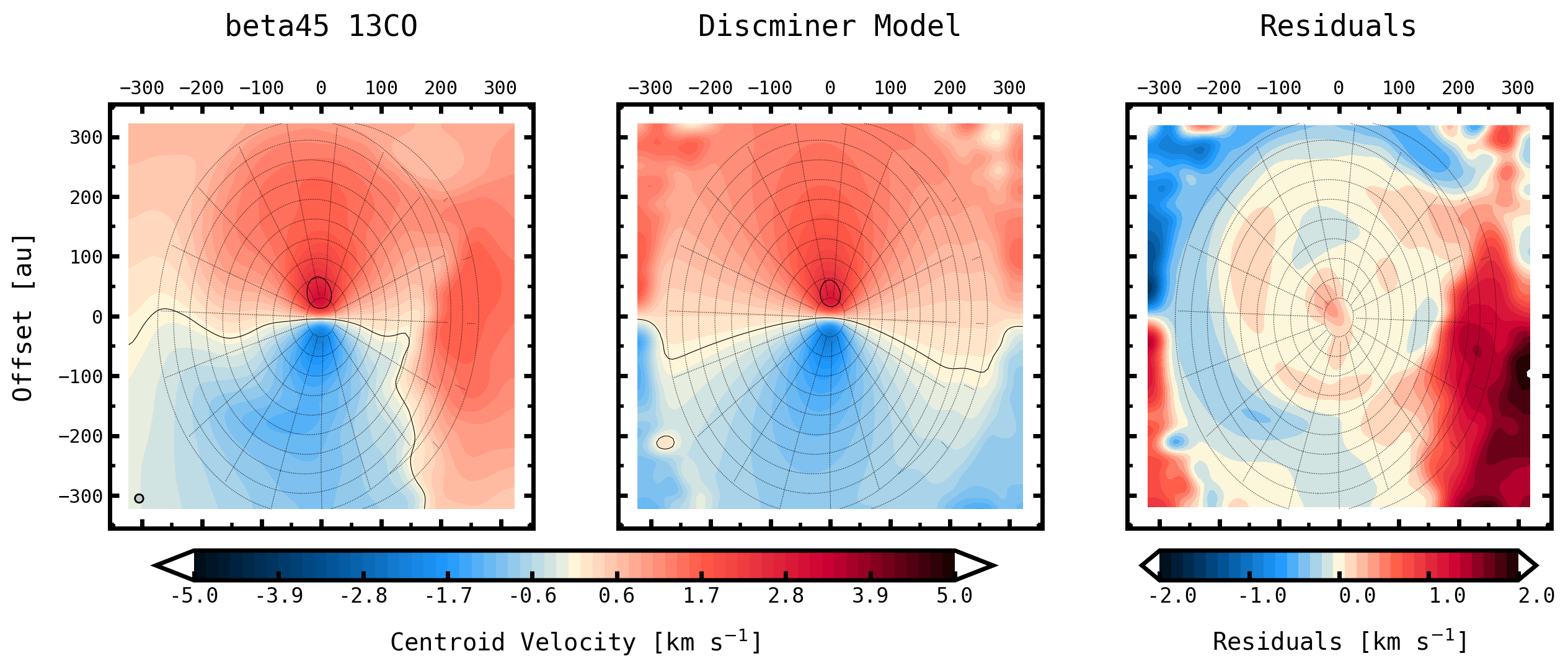}
        \caption{Velocity moment map and residuals for the \textsuperscript{13}CO emission.}
        \label{fig:13CO_residuals}
    \end{subfigure}
    \caption{Left panel: LOS centroid velocity for the synthetic ALMA observation of the \texttt{beta45} simulation 5402 yr after periastron, middle panel: \textsc{discminer} model for a smooth Keplerian disc, and right panel: residual LOS velocity of the synthetic observation when a smooth Keplerian model is subtracted off, for (\subref{fig:CO_residuals}) the CO and (\subref{fig:13CO_residuals}) the \textsuperscript{13}CO emission.}
    \label{fig:velocity_moment+residuals}
\end{figure*}

\autoref{fig:velocity_moment+residuals} shows the centroid velocity of the synthetic observations and the smooth Keplerian model as well as the residual of the two for both the CO and \textsuperscript{13}CO molecular line emission in the \texttt{beta45} simulation. We observe spiral-like structures and azimuthal asymmetries in the residuals of both molecules. There is little difference between the two tracers, with red-shifted and blue-shifted features in the velocity residuals at the same locations and of roughly the same magnitudes and morphologies. This is despite the fact that CO and \textsuperscript{13}CO trace different heights in the disc's emitting surface; we can hence conclude that the perturbations to the velocity field induced by the flyby do not strongly depend on height.

\begin{figure*}
    \centering
    \begin{subfigure}{\textwidth}
        \includegraphics{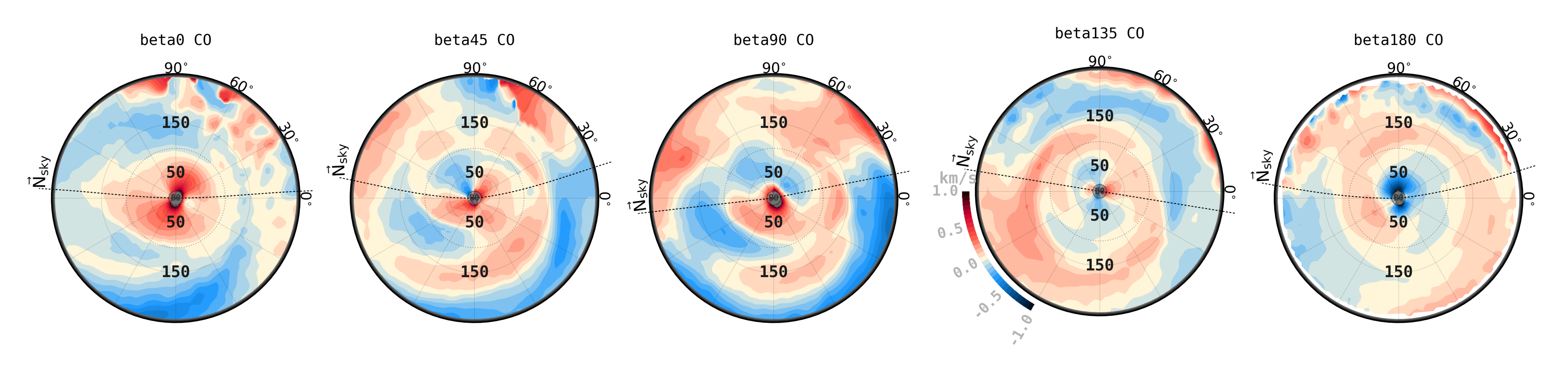}
        \caption{Velocity residuals for the CO emission.}
        \label{fig:CO_deproj_residuals}
    \end{subfigure}
    \begin{subfigure}{\textwidth}
    \includegraphics{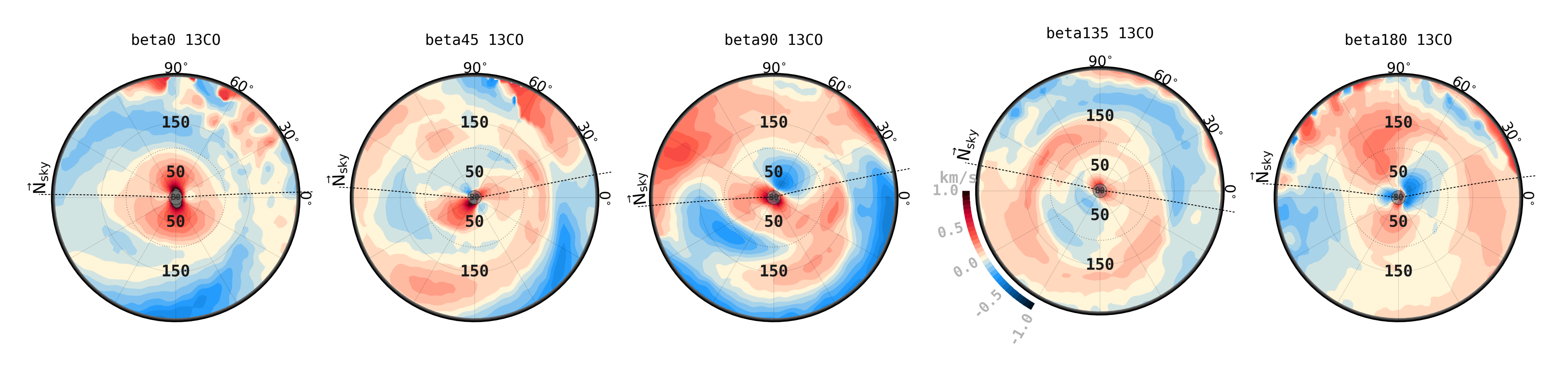}
        \caption{Velocity residuals for the \textsuperscript{13}CO emission.}
        \label{fig:13CO_deproj_residuals}
    \end{subfigure}
    \caption{Centroid velocity residuals for the (\subref{fig:CO_deproj_residuals}) CO and (\subref{fig:13CO_deproj_residuals}) \textsuperscript{13}CO emission from (left to right): the \texttt{beta0}, \texttt{beta45}, \texttt{beta90}, \texttt{beta135} and \texttt{beta180} simulations 5402 yr after periastron, deprojected into the reference frame of the disc. Note that colorbars are the same for all figures but only shown on the \texttt{beta135} residual maps.}
    \label{fig:deproj_residuals}
\end{figure*}

\autoref{fig:deproj_residuals} shows the velocity residuals in the CO and the \textsuperscript{13}CO for all five simulations, now deprojected from the sky plane into the plane of the disc itself. We see elongated features resembling spirals in the velocity residuals in all simulations but \texttt{beta0}, which also shows some non-Keplerian features in the CO gas kinematics, but these substructures are more axisymmetric than those seen in the other simulations. This may be due to the more efficient truncation of the disc in the \texttt{beta0} encounter, as well as the fact that the density spirals dissipate soonest in a prograde coplanar flyby, as seen in \autoref{fig:faceon_densities}. In the three inclined flyby simulations, as well as the coplanar retrograde flyby \texttt{beta180}, spirals are visible in the velocity residuals, despite the disc \emph{not} being warped in the case of a coplanar flyby. We again see little difference between the velocity residuals for the CO and \textsuperscript{13}CO emission.

\begin{figure*}
    \centering
    \begin{subfigure}{\textwidth}
        \includegraphics[scale=0.4]{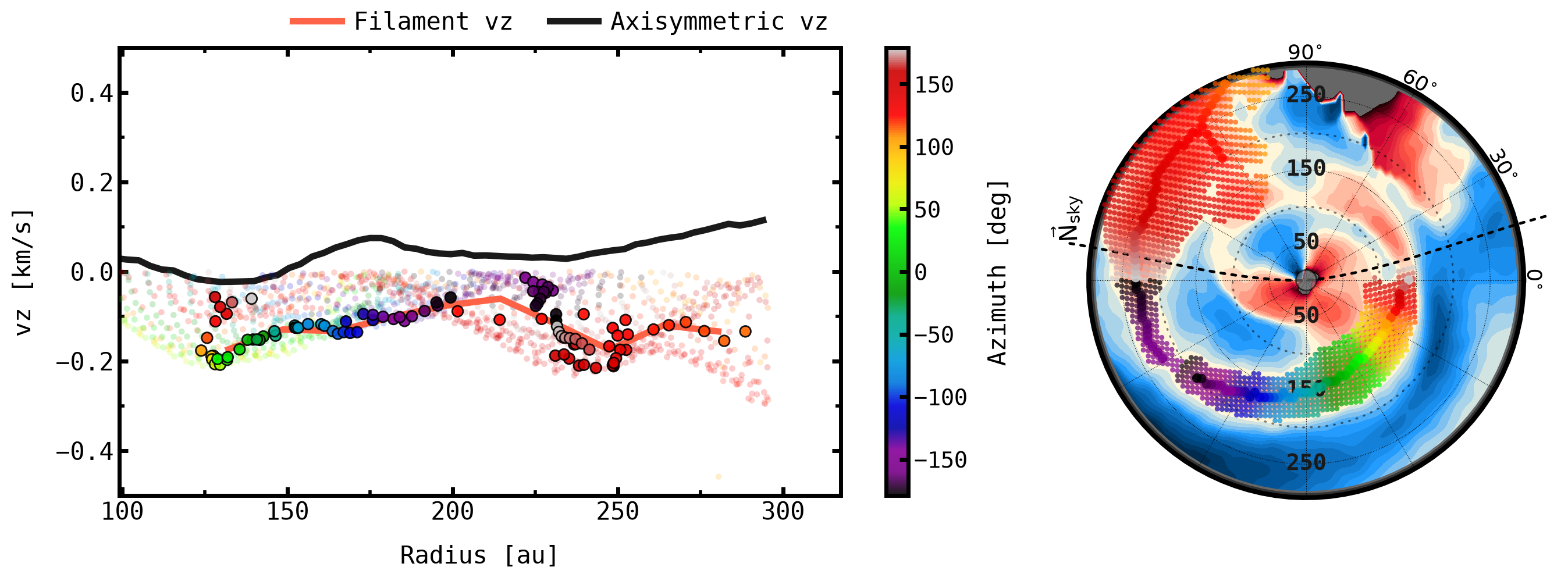}
        \caption{\texttt{beta45}}
        \label{fig:beta45_CO_filaments}
    \end{subfigure}
    \begin{subfigure}{\textwidth}
        \includegraphics[scale=0.4]{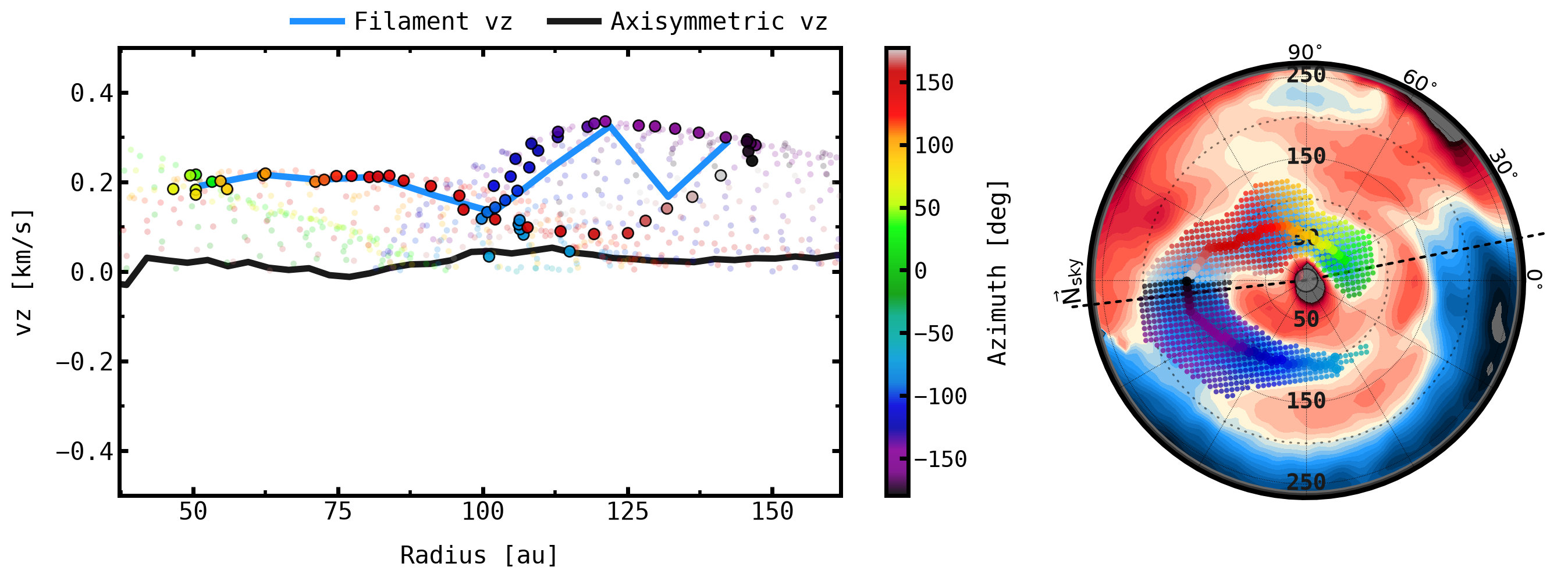}
        \caption{\texttt{beta90}}
        \label{fig:beta90_CO_filaments}
    \end{subfigure}
    \begin{subfigure}{\textwidth}
        \includegraphics[scale=0.4]{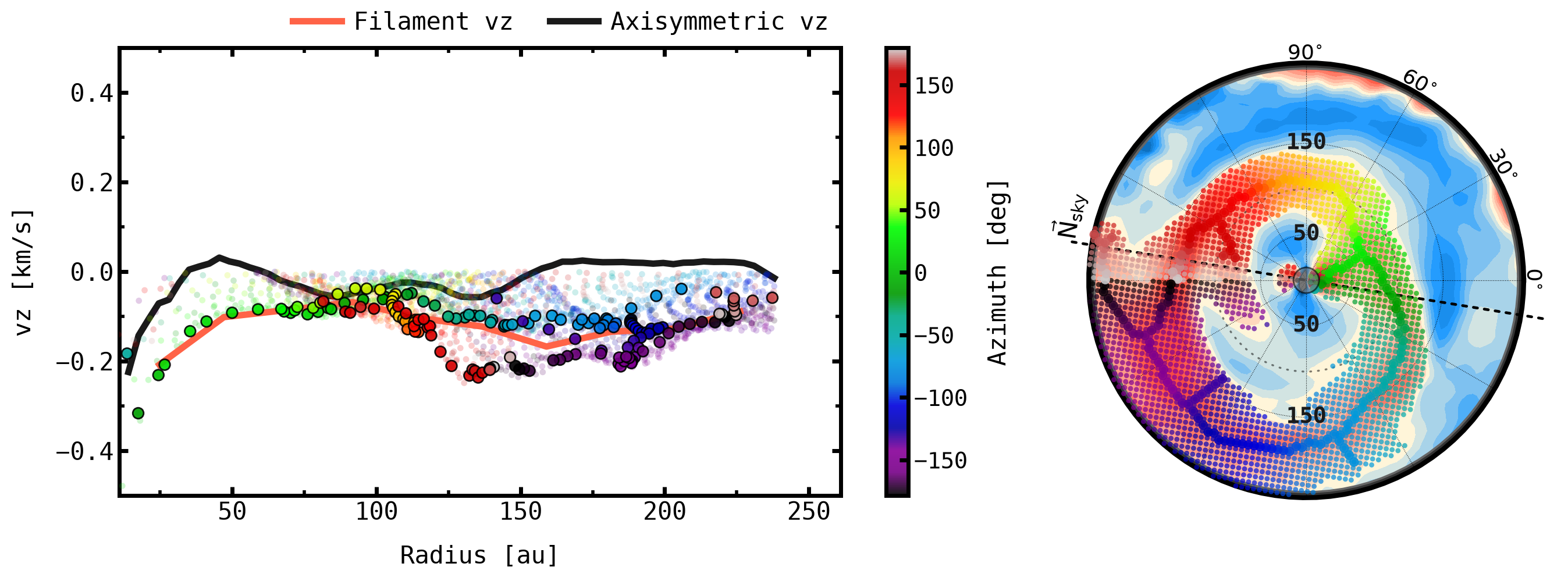}
        \caption{\texttt{beta135}}
        \label{fig:beta135_CO_filaments}
    \end{subfigure}
    \begin{subfigure}{\textwidth}
        \includegraphics[scale=0.4]{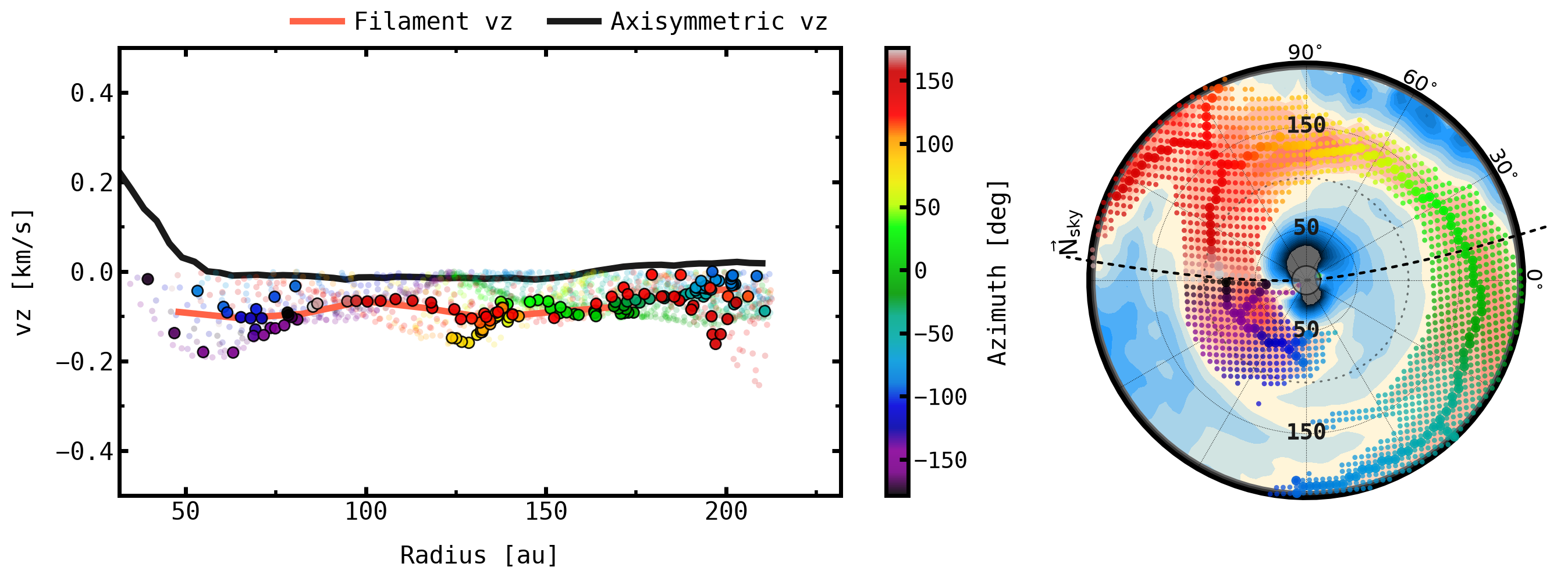}
        \caption{\texttt{beta180}}
        \label{fig:beta180_CO_filaments}
    \end{subfigure}
    \caption{Line of sight velocities measured along a filament for the CO emission 5402 yr after periastron, for the (\subref{fig:beta45_CO_filaments}) \texttt{beta45}, (\subref{fig:beta90_CO_filaments}) \texttt{beta90}, (\subref{fig:beta135_CO_filaments}) \texttt{beta135} and (\subref{fig:beta180_CO_filaments}) \texttt{beta180} simulations. Dark coloured points are velocities extracted along the spine of the filaments depicted in the velocity residual map in the right panels, while pale dots trace velocities measured along the width of the filaments.}
    \label{fig:velocity_filaments_all}
\end{figure*}

We trace the spiral features in the velocity residuals and determine the peak LOS velocities along the spine of the spiral features. Following the procedure outlined by \citet{Benisty_2026}, we trace the non-Keplerian substructures in the velocity residuals using \textsc{FilFinder} \citep{Koch_2015}. The results of this process for the CO emission for all simulations (except the \texttt{beta0} simulation, which does not have a clear spiral structure) can be seen in \autoref{fig:velocity_filaments_all}. In the \texttt{beta45} and \texttt{beta90} simulations, there are prominent spiral arms with a spiral arm with a maximum blue- and red-shifted velocities of up to 300 m/s. However, in both retrograde flybys, \texttt{beta135} and \texttt{beta180}, the filaments in the velocity residuals have smaller peak velocities, with magnitudes of not more than 100 m/s.

\subsection{Fitting warp profiles}

The azimuthal asymmetries in the disc frame can now be fit to the predicted velocity residuals caused by warping in the disc using \autoref{eqn:vlos_warp} in order to extract out the disc's warp profile in the sky plane, $\delta i(R)$ and $\delta\mathrm{PA}(R)$, and hence the disc-frame quantities $\beta(R)$ and $\gamma(R)$. We use the \texttt{python} package \texttt{warpfitter} \citep{Winter_2025} to perform these fits and extract radial profiles for $\beta(R)$ and $\gamma(R)$ \emph{assuming that the velocity residuals are entirely caused by the geometry of the warped disc}. Full results from the warp fitting procedure are presented in Appendix \ref{sec:warpfitter_extra}.

\begin{figure*}
    \centering
    \begin{subfigure}{\textwidth}
        \includegraphics{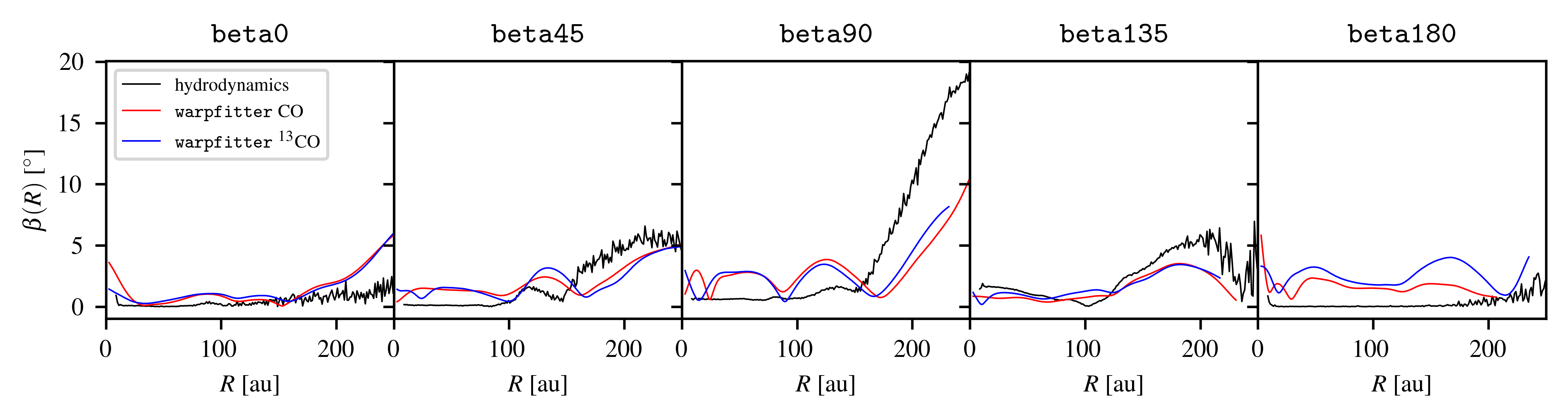}
        \caption{Radial profiles of the tilt angle $\beta(R)$.}
        \label{fig:beta_warpfitter}
    \end{subfigure}
    \begin{subfigure}{\textwidth}
        \includegraphics{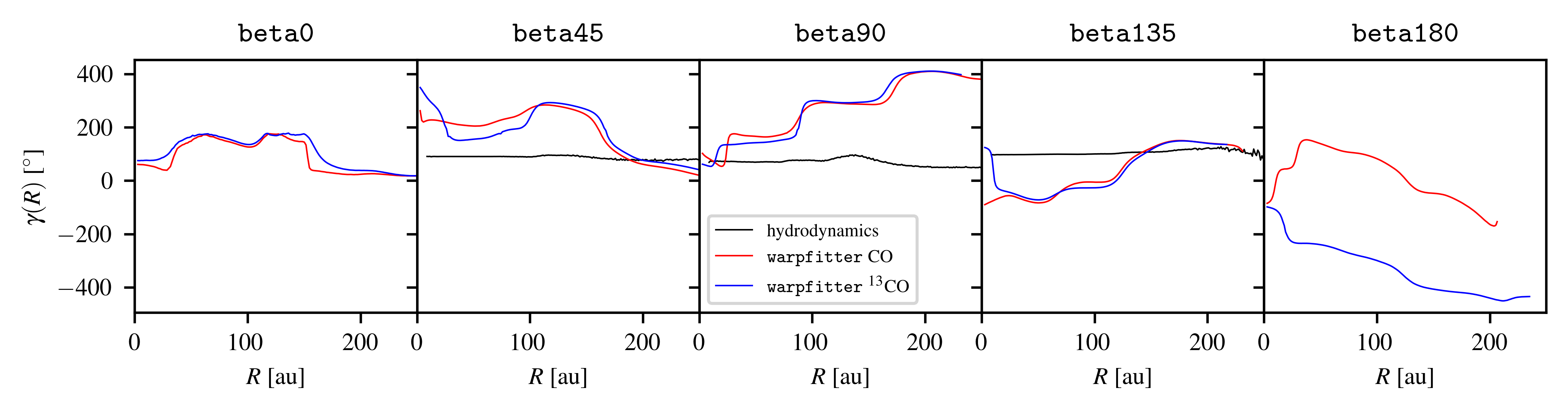}
        \caption{Radial profiles for the twist angle $\gamma(R)$.}
        \label{fig:gamma_warpfitter}
    \end{subfigure}
    \begin{subfigure}{\textwidth}
        \includegraphics{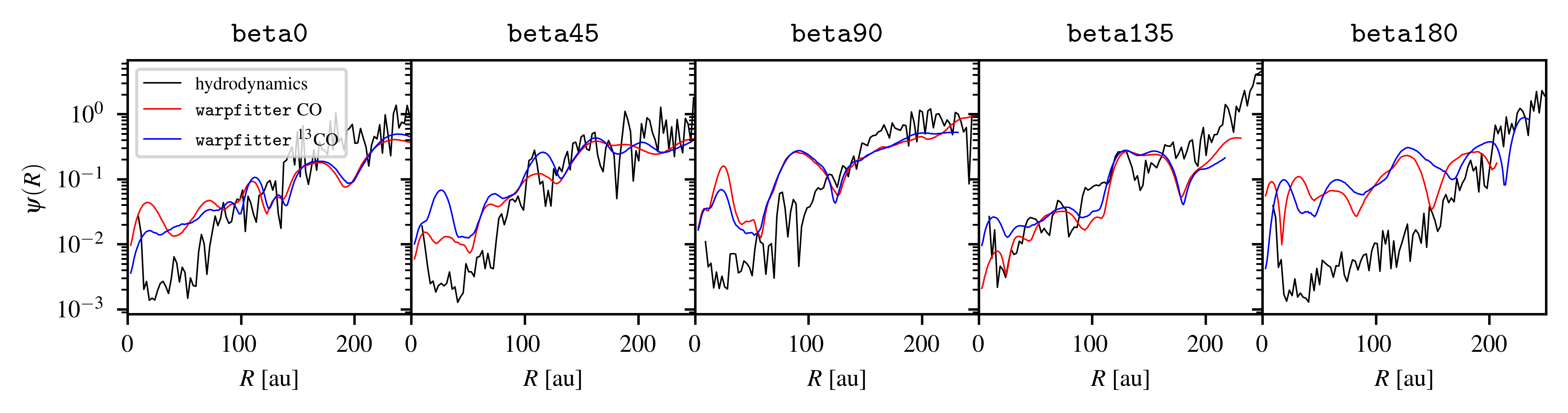}
        \caption{Radial profiles for the dimensionless warp amplitude $\psi(R)$.}
        \label{fig:psi_warpfitter}
    \end{subfigure}
    \caption{Radial profiles of (\subref{fig:beta_warpfitter}) the tilt angle $\beta(R)$, (\subref{fig:gamma_warpfitter}) the twist angle $\gamma(R)$ and (\subref{fig:psi_warpfitter}) the warp amplitude $\psi(R)$ 5402 yr after periastron for (left to right): the \texttt{beta0}, \texttt{beta45}, \texttt{beta90}, \texttt{beta135} and \texttt{beta180} simulations. Black lines show the profiles extracted from \textsc{Phantom} output. Note that the hydrodynamical profiles for $\beta(R)$ are shown in a coordinate frame rotated by the global inclination of the disc. Note also that as the twist angle is ill-defined for a non-warped disc, no hydrodynamical profiles for $\gamma(R)$ are shown in the two coplanar flyby simulations \texttt{beta0} and \texttt{beta180}. Note further that, unlike in Figures \ref{fig:beta_evolution} and \ref{fig:gamma_evolution}, the radial profiles extend to 250 AU, beyond the initial extent of the unperturbed disc, and we expect noisier hydrodynamical profiles in these outer regions where there is little disc material. Red and blue lines show, respectively, the profiles extracted by \texttt{warpfitter} for the CO and \textsuperscript{13}CO emission from the synthetic observations.}
    \label{fig:warpfitter_all}
\end{figure*}

\autoref{fig:warpfitter_all} shows the results of this process for all five flyby encounters. (Note that for completeness, we show the results of fitting the features in the velocity residuals to a warp for the \texttt{beta0} simulation, although we know that there is no warping in the coplanar simulations and \autoref{fig:deproj_residuals} shows that there are no spiral features in the velocity residuals in the \texttt{beta0} case.) 

In \autoref{fig:beta_warpfitter}, we compare the profiles extracted by \texttt{warpfitter} for the tilt angle $\beta(R)$ to the tilt angle computed directly from the hydrodynamics. Noting that the three inclined flybys impart total inclinations of between 5 and $15^\degree$  to the disc by the end of the simulation, as can be seen in \autoref{fig:beta_evolution}, we choose to show the hydrodynamical $\beta(R)$ profiles in a Cartesian coordinate frame \emph{rotated by the global inclination of the disc}, as discussed in \autoref{sec:warped_coords} and shown as the dashed line in \autoref{fig:warp_schematic} and in the middle panel of \autoref{fig:vz_comparison}. This enables us to better compare the profiles extracted from the hydrodynamical results to those computed by \texttt{warpfitter}, as the \texttt{warpfitter} input is synthetic observations and it does not know the ``true'' inclination of the disc in its own frame. 

As in \autoref{fig:beta_evolution}, we see that there is little radial variation in the tilt angle profile $\beta(R)$ of the disc by the end of the simulation. We see moderate warping of a few degrees in the inclined flyby encounters, particularly the \texttt{beta90} simulation, and much smaller misalignments in the two coplanar encounters, \texttt{beta0} and \texttt{beta180}. The CO and \textsuperscript{13}CO emission from the synthetic observations agree quite closely with the hydrodynamical tilt profiles in most cases, although they vary more strongly with radius and hence imply the discs are more warped than can be seen in the hydrodynamical profiles. For the coplanar cases \texttt{beta0} and \texttt{beta180}, the tilt is consistently overestimated by a few degrees compared to the hydrodynamical value. We also note that in the \texttt{beta90} case, the tilt angle profile in the outer region of the disc ($R>150$ au) is \emph{under}estimated compared to the hydrodynamical tilt profile; we can see from \autoref{fig:edgeon_densities} that there is little material in the outermost part of the disc, and so it is likely that the synthetic observations in this region are noisier and harder to fit to a simple warp profile.

\autoref{fig:gamma_warpfitter} shows the equivalent comparison for the hydrodynamical and \texttt{warpfitter} \emph{twist} angle profiles. Here, we find dramatic differences between the hydrodynamical $\gamma(R)$ profile and those predicted by \texttt{warpfitter}. Although \autoref{fig:gamma_evolution} suggests that the flyby-induced twisting in the disc does \emph{not} persist long after the flyby, \autoref{fig:gamma_warpfitter} shows significant variation in the $\gamma(R)$ profiles for all five simulations, even the two coplanar encounters where the disc is not warped at all and the twist is ill-defined. (As a caveat, we note that the value of $\gamma$ is ill-defined when the disc is not warped and $\beta\approx 0$. Some of the large variations in the \texttt{warpfitter} $\gamma(R)$ profiles correspond to radii where the value of $\beta(R)$ is at a local minimum close to 0, most notable in the \texttt{beta90} case, and so the discontinuities in the $\gamma(R)$ profiles at these locations may be largely numerical instead of corresponding to a highly twisted disc. However, we note that the \texttt{warpfitter} $\gamma(R)$ profiles are also highly varying in many locations where the tilt angle is not small, and in places whether the corresponding hydrodynamical profile of $\gamma(R)$ is constant with radius.)

\autoref{fig:psi_warpfitter} quantifies the warp amplitude $\psi(R)$ of the disc to show that, although \texttt{warpfitter} significantly overestimates the twist angle in the disc, its predictions for the amplitude of warping in the disc is fairly similar to the warp amplitude extracted from the hydrodynamics in the three inclined flyby cases. We attribute this to the fact that the warp amplitude is computed from the extracted \texttt{warpfitter} profiles of \emph{both} $\beta(R)$ and $\gamma(R)$, and although the \texttt{warpfitter} $\gamma(R)$ profiles differ significantly from the hydrodynamics, the $\beta(R)$ profiles do not. However, in the coplanar flyby cases, the warp amplitude is overestimated by about an order of magnitude in the inner regions of the disc.

It therefore follows that the kinematic substructures observed in a flyby-perturbed disc \emph{cannot} all be attributed to warping in the disc, and fitting the velocity residuals to a prescribed warp using \autoref{eqn:vlos_warp} will result in significantly overestimating the level of twisting in the disc. As could be inferred from \autoref{fig:vz_comparison}, flybys induce kinematic substructures in discs beyond those caused by the geometry of warping.

\subsection{Multi-wavelength observational signatures}

We have demonstrated that the kinematic signatures induced by flybys may be erroneously interpreted as warps. It is also important to consider other observational signatures of flybys in order to determine whether or not they can be uniquely identified. 

\begin{figure*}
    \centering
    \includegraphics{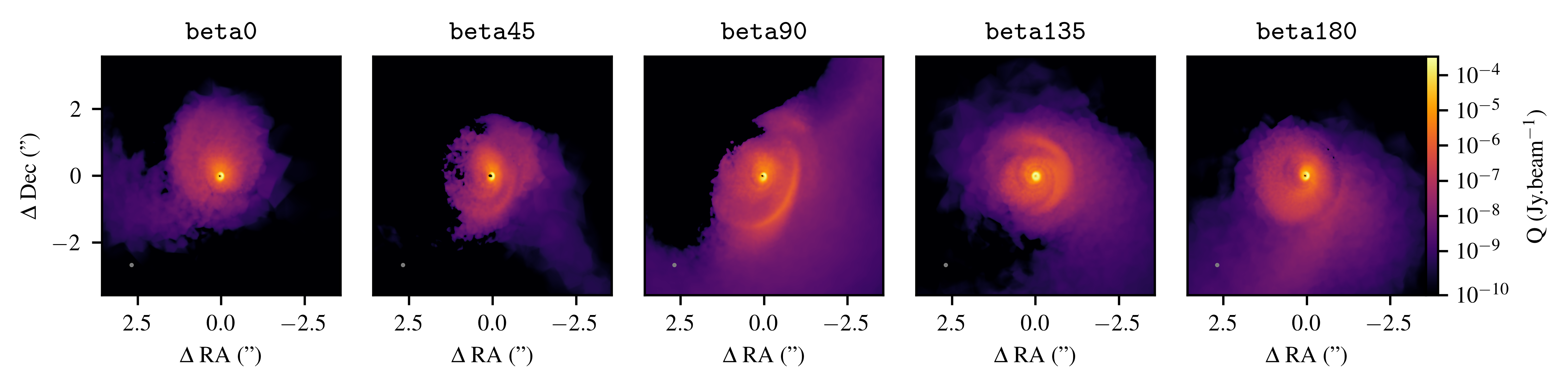}
    \caption{\texttt{pymcfost} synthetic VLT/SPHERE observations of scattered light emission at a wavelength of $1.65\mu\mathrm{m}$,  for (left to right): the \texttt{beta0}, \texttt{beta45}, \texttt{beta90}, \texttt{beta135} and \texttt{beta180} simulations.}
    \label{fig:scattered_light}
\end{figure*}

\begin{figure*}
    \centering
    \includegraphics{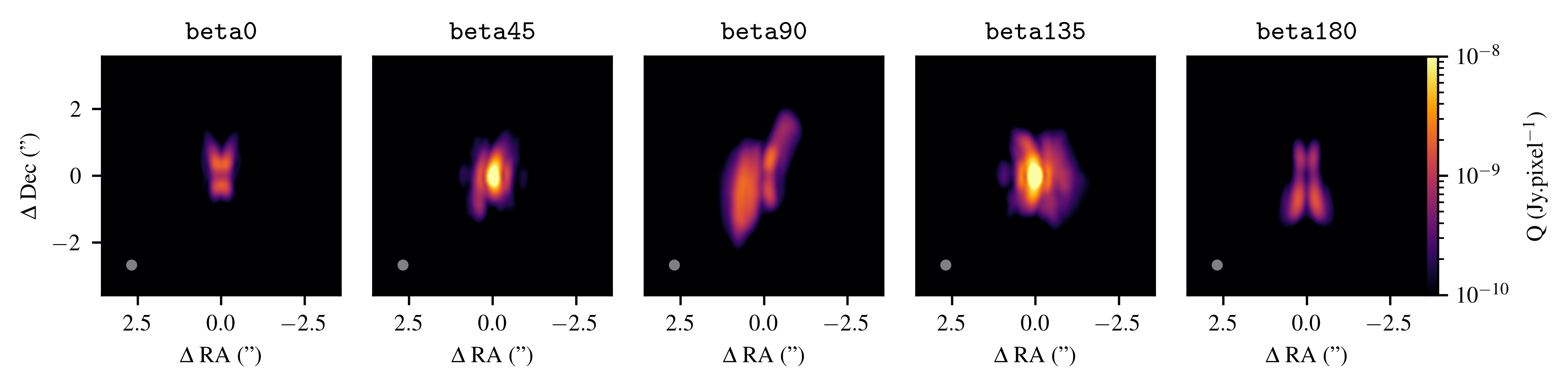}
    \caption{\texttt{pymcfost} synthetic JWST observations of edge-on scattered light emission at a wavelength of $7.7\mu\mathrm{m}$,  for (left to right): the \texttt{beta0}, \texttt{beta45}, \texttt{beta90}, \texttt{beta135} and \texttt{beta180} simulations.}
    \label{fig:scattered_light_edgeon}
\end{figure*}

\begin{figure*}
    \centering
    \includegraphics{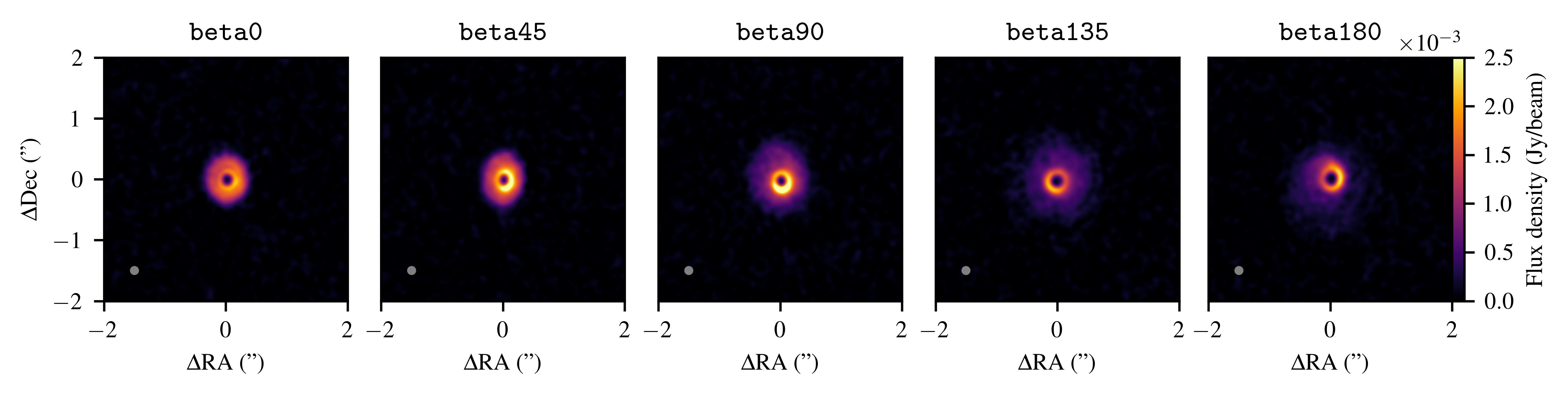}
    \caption{\texttt{pymcfost} synthetic ALMA observations of dust continuum emission at a wavelength of 1.3mm,  for (left to right): the \texttt{beta0}, \texttt{beta45}, \texttt{beta90}, \texttt{beta135} and \texttt{beta180} simulations.}
    \label{fig:continuum}
\end{figure*}

\autoref{fig:scattered_light} shows synthetic VLT/SPHERE scattered light observations of the disc 5402 yr after the flyby for all five simulations, at a wavelength of 1.65 $\mu\mathrm{m}$. In the three inclined flyby simulations \texttt{beta45}, \texttt{beta90}, \texttt{beta135}, spiral features can be seen in the disc; however, no spiral arms are visible in the coplanar simulations \texttt{beta0} and \texttt{beta180}. 

\autoref{fig:scattered_light_edgeon} shows synthetic JWST scattered light observations of the disc in the edge-on configuration, i.e. with radiative transfer calculations performed assuming that the disc is inclined at $90\degree$ to the line of sight of the observer, rather than $30\degree$ as used in all other radiative transfer calculations presented in this work. In the edge-on configuration, discs appear as two nebulae (corresponding to the two emitting surfaces of the disc) separated by a dark line (corresponding to the optically thick midplane of the disc). In the three inclined flyby simulations \texttt{beta45}, \texttt{beta90}, \texttt{beta135}, we see a lateral asymmetry between the two nebulae, with one nebulae misaligned relative to the other nebulae. This lateral asymmetry is not visible in the two coplanar simulations. These results agree with previous studies suggesting that lateral asymmetry in scattered light observations of discs viewed edge-on are caused by misalignments between the discs' inner and outer regions, and can be observed even when the misalignments are on the order of only a few degrees \citep{Villenave_2024,Kimmig_Villenave_2025}.

\autoref{fig:continuum} shows corresponding synthetic ALMA continuum emission observations of the disc at a wavelength of 1.3mm; no substructure can be seen in the mm-sized dust grains with ALMA. We can therefore conclude that flyby-induced spirals can \emph{not} be expected to be observed in ALMA continuum emission, but may be visible in the scattered light observations. We note that continuum emission traces larger (mm-sized) dust grains that we expect to have settled to the disc midplane in a planar disc, whereas scattered light observations trace smaller ($\mu\mathrm{m}$-sized) dust grains in the upper layers of the disc that are well-coupled to the gas. The perturbations induced by flybys therefore appear to be more easily observed in the upper layers of the disc than the disc midplane. We note that that several observed candidates for flyby-perturbed discs appear axisymmetric in continuum emission, such as RW Aur \citep{Kurtovic_2024} and UX Tau A \citep{Zapata_2020}.

\section{Discussion} \label{sec:discussion}

\subsection{Velocity residuals and kinematic features}

We have seen that stellar flybys induce substructures in the disc's  velocity that are visible even when the warp induced by the disc is accounted for fully, as in \autoref{fig:vz_comparison}. These kinematic features must therefore be caused by the interaction of the disc with the perturbing star, rather than being simple geometric features of the warp itself. We note that there is significant complexity in the disc kinematics long after the flyby-induced spiral features in the disc density are no longer qualitatively visible, suggesting that signatures of stellar flybys are observable in kinematics for a long time after the periastron of the flyby itself.

For all flyby encounters except the coplanar prograde flyby \texttt{beta0}, we find spiral features in the residual LOS velocity of the disc compared to a smooth Keplerian disc several thousand years after the flyby periastron. The spirals have maximum LOS velocities of about 200 m/s (for inclined prograde and coplanar flybys) and 100 m/s (for retrograde flybys). The finding of spirals in the CO velocity residuals is in agreement with the results of \citet[figure 16]{Kimmig_2026}, who model the RW Aur system as an inclined flyby. \citeauthor{Kimmig_2026} suggest that the spirals in the velocity residuals may originate from the spirals triggered by the flyby itself or from the warped morphology of the disc; our results in \autoref{fig:warpfitter_all} show that spirals are present even in the coplanar flyby simulations when the disc is not warped, indicating the most likely origin for the spiral features is the flyby itself and not the warp it induces on inclined orbits. 

Attempting to fit the spiral features in the velocity residuals results in twist angle $\gamma(R)$ profiles that vary significantly with radius, although the hydrodynamical results show that the twists induced by inclined flybys have dissipated by several thousand years after the flyby. We see in \autoref{fig:psi_warpfitter} that assuming all spirals in the velocity residuals are caused by warping in the disc results in significantly over-predicting the warp amplitude; this remain the case in the two coplanar flyby simulations, \texttt{beta0} and \texttt{beta180}, which are unable to warp the disc. It is therefore clear that both coplanar and inclined flybys imprint spiral features in the disc's CO and \textsuperscript{13}CO kinematics, regardless of whether or not a warp is present. These spiral features can be attributed to the tidal perturbation caused by the flyby; as seen in \autoref{fig:density_vz_comparison}, they appear to persist for longer in the disc than the spirals seen in the gas density. The kinematic imprints of stellar flybys on protoplanetary discs are hence longer-lived than the effects on the disc density.

\subsection{Warps induced by inclined flybys} \label{sec:warp_discussion}

We note that, similarly to \citet{Kimmig_2026}, we find that inclined flybys excite warps in the disc and coplanar flybys do not. The results of our hydrodynamical models show that the warping in the disc is damped by 5402 yr after the flyby (approximately 3 orbital times at the disc's outer edge, $\Omega_\mathrm{out}^{-1}$). \citeauthor{Kimmig_2026} observe that a prograde inclined flyby with the point of periastron in the same plane as the disc (analogous to the flyby orbits presented in this work) is warped for about $11 \Omega_\mathrm{out}^{-1}$ after the flyby. They also study inclined flybys whose orbits are rotated so that the periastron is not in the same plane as the disc, and find that warps can persist for as long as $100 \Omega_\mathrm{out}^{-1}$. We attribute this difference in warp survival times between the two works to the different hydrodynamical methods used; the higher viscosity used in SPH causes stronger damping of the warp than the grid-based simulations employed by \citeauthor{Kimmig_2026}. We underline that, since our SPH simulations use a higher viscosity that cause flyby-induced warps to dissipate faster than expected, the results presented in this work roughly correspond to those of the more physical low-viscosity case at later times (assuming the disc remains in the same warp regime in both the higher and lower viscosity cases). We also note that \citeauthor{Kimmig_2026} found the strongest and longest-lasting warps were induced by inclined flyby encounters with the periastron out of the plane of the disc; these encounters are statistically more likely, but for simplicity we only simulated flybys with periastron in the same plane as the disc.

For this reason, it is beyond the scope of this work to study the evolution of the warp via shear flows, known as sloshing and breathing modes (see \citealt{Dullemond_2022}). As these shear motions and the parametric instability that they generate in warped discs \citep{Ogilvie_Latter_2013} are both inversely proportional to the viscosity in the disc, we do not expect to recover either in high-viscosity SPH simulations. Sloshing and breathing modes are also strongest on disc surfaces, where the gas density is low and the SPH resolution is hence also too low to recover them.

\subsection{Relevance to exoALMA results}

The simple warp model presented by \citet{Winter_2025}  (\autoref{eqn:vlos_warp}) results in an $m=1$ spiral in the velocity residuals. \citeauthor{Winter_2025} observe these features most strongly in the exoALMA targets MWC 758 and CQ Tau. However, the velocity residuals for our synthetic observations, shown in \autoref{fig:deproj_residuals}, show more azimuthal asymmetry than can be described by a clean $m=1$ spiral, supporting our claim that flybys induce more kinematic complexity than can be simply ascribed to a warp.

We note that the residual velocity fields for the two retrograde flyby simulations, \texttt{beta135} and \texttt{beta180}, show the clearest morphological resemblance to the observed residual velocity field of MWC 758. In Figures \ref{fig:beta135_co_warpfitter} and \ref{fig:beta135_13co_warpfitter}, we also see some similarities between the warp profiles inferred for the disc in the \texttt{beta135} simulation and for MWC 758 \citep[see][figures 1 and 2]{Winter_2025}. Most strikingly, we see almost linear (periodic) $\gamma(R)$ profiles in both MWC 758 and the \texttt{beta135} inferred \texttt{warpfitter} results, which \citeauthor{Winter_2025} suggest is the cause of the clear spiral structure in the disc's velocity residuals. However, \autoref{fig:gamma_evolution} shows that the true value of $\gamma(R)$ for the \texttt{beta135} disc 5402 yr after periastron is in fact nearly constant at about $90\degree$ -- so the spirals visible in the velocity residuals cannot have arisen as a result of the twist profile.

We also note the striking similarity between the \texttt{warpfitter} results for the \texttt{beta45} simulation (Figures \ref{fig:beta45_co_warpfitter} and \ref{fig:beta45_13co_warpfitter}) and the disc CQ Tau \citep[see][figure C4]{Winter_2025}. An $m=1$ spiral structure can be seen in the velocity residuals CQ Tau's CO emission (although interestingly \emph{not} the \textsuperscript{13}CO emission) that is very similar to the redshifted structure visible in the velocity residuals of the \texttt{beta45} synthetic observations in \autoref{fig:deproj_residuals}. For both CQ Tau and the \texttt{beta45} synthetic observations, \texttt{warpfitter} infers a twist angle profile that (periodically) decreases linearly with radius, although once again \autoref{fig:gamma_evolution} shows that the twist angle is actually nearly constant with radius 5402 yr after periastron in the \texttt{beta45} simulation. There is strong evidence from VLTI/GRAVITY and ALMA observations that CQ Tau is in fact a misaligned disc \citep{Bohn_2022, Ishibashi_2026}, and it has been suggested that the misalignment may be caused by an unresolved close-in companion on an inclined, eccentric orbit \citep{Hammond_2022}. We therefore do \emph{not} suggest that the spirals visible in CQ Tau's CO velocity residuals are caused by a past stellar flyby. Instead, we note the similarity between the effect of an inclined prograde stellar flyby, such as that in the \texttt{beta45} simulation, and that of a lower-mass internal companion on an inclined orbit on the disc kinematics.

\citet{Benisty_2026} study the velocities of the spiral structures observed in the velocity residuals of MWC 758 and CQ Tau. They find that the large redshifted spiral in MWC 758 has a peak downward velocity of -265 m/s. We note that the similar redshifted spirals in the retrograde flyby simulations \texttt{beta135} and \texttt{beta180} have much lower peak downward velocities of only about -100 m/s, as shown in Figures \ref{fig:beta135_CO_filaments} and \ref{fig:beta180_CO_filaments}. \citeauthor{Benisty_2026} also find that the redshifted spiral structure seen in CQ Tau has a peak downward velocity of -290 m/s; we see in \autoref{fig:beta45_CO_filaments} that the redshifted spiral in the \texttt{beta45} velocity residuals has a peak downward velocity of no more than 200 m/s. We note that stellar flybys do not seem capable of producing LOS velocity perturbations as large as those observed in these two exoALMA targets several thousand years after the flyby periastron.

\subsection{Signatures of flybys in scattered light emission}

Figures \ref{fig:scattered_light}, \ref{fig:scattered_light_edgeon} and \ref{fig:continuum} demonstrate that flyby-induced substructures may be observable in the scattered light emission, but not in the dust continuum emission. We see in \autoref{fig:scattered_light} that spirals are visible in the three inclined flybys \texttt{beta45}, \texttt{beta90} and \texttt{beta135}, but no substructure can be seen in the two coplanar cases.

The presence of spiral arms only in the inclined flyby cases, where the disc is slightly warped as a result of the encounter, suggests that the spirals in scattered light are caused by the presence of the warp itself. These results are in agreement with the modelling conducted by \citet{Winter_2025}, who are able to reproduce some of the spiral structure seen in the scattered light emission of the disc MWC 758 with a simple warp profile. 

These signatures in the scattered light could be used to distinguish between a disc that is warped and one that has additional flyby-induced perturbations. We have shown that flyby-induced spirals in the disc kinematics can be mistaken for warps, but the presence of a spiral in the scattered light allows us to confirm that the disc is in fact warped, whereas the lack of a scattered light spiral suggests that the disc is unwarped and the kinematic features are due to a near-coplanar flyby. It is even possible that the morphology of the scattered light spiral could be used to constrain the orbit of a potential recent stellar flyby, although this study is beyond the scope of this work.

Additionally, \autoref{fig:scattered_light_edgeon} shows that the inclined flyby cases show lateral asymmetry between the disc's two nebulae, with one nebular more extended to the south than the other. \citet{Villenave_2024} identify this phenomenon in 15 edge-on discs observed with scattered light and attribute lateral asymmetries to misalignments between the inner and outer regions of the discs. \citet{Kimmig_Villenave_2025} performed radiative transfer simulations of edge-on warped protoplanetary discs to determine the signatures of warps in scattered light emission. They found that, for optimal azimuthal viewing angles, misalignments as low as $2\degree$ could result in visible lateral asymmetries in the scattered light observations, and that larger misalignments resulted in stronger lateral asymmetries. As our inclined flyby encounters produce misalignments between the inner and outer discs of between $5\degree$ and as much as $20\degree$, as can be seen in \autoref{fig:beta_warpfitter}, we attribute the lateral asymmetry seen in \autoref{fig:scattered_light_edgeon} to warps induced by inclined flybys. Together, these results show that the warps induced by inclined flybys appear to be visible in scattered light emission for a variety of disc inclinations: spirals can be seen in warped discs in near-face-on configurations, and lateral asymmetries when the disc is viewed edge-on. 

Recent work by \citet{Milsom_2026} has demonstrated that the warps induced by inclined flybys can be observed via long-lasting shadows in the disc's scattered-light emission. Qualitatively, we do not find any such shadowing in our synthetic scattered light observations. We note that \citeauthor{Milsom_2026} stress that very small warps will not leave observable shadows in the scattered light, even when they are observable in the disc's kinematics. It is therefore possible that the relatively small degree of flyby-induced warping seen long after periastron in our simulations (exacerbated by the faster dissipation of warps in high-viscosity SPH simulations) is not observable as shadows in the scattered light emission.

\section{Conclusions} \label{sec:conclusions}

We have studied the response of a protoplanetary disc to inclined and coplanar stellar flybys using 3D SPH simulations. Our results agree with previous findings that inclined stellar flybys whose periastron is in the plane of the disc can induce spirals and warps in the disc; however, by a few thousand years after the moment of periastron, the disc is only very slightly warped and the twist induced by the flyby has dissipated, leaving an untwisted disc. We also demonstrate that the warping of the disc must be considered in order to fully separate the vertical motions in the disc from the component due to the Keplerian rotation of the disc. 

We have used radiative transfer modelling to create synthetic observations of discs perturbed by flybys and studied the velocity residuals in the discs' CO and \textsuperscript{13}CO emission compared to a smooth Keplerian disc. We find that retrograde and inclined coplanar flybys leave large-scale non-Keplerian features in the disc kinematics.

When we attempt to fit these features in the velocity residuals to a simple analytical warp profile, we we find that the best-fit warp profile predicts a disc that is significantly more twisted than can be seen in the hydrodynamics. We therefore determine that flybys introduce significant non-Keplerian structure in the disc kinematics above and beyond those that can be explained by warping of the disc. We note that attempting to attribute non-Keplerian features in the velocity residuals to warps alone may result in over-predicting the level of warping in a disc that was in fact perturbed by a stellar flyby.

We also present synthetic observations of scattered light and thermal continuum emission from discs perturbed by inclined and coplanar stellar flybys. We find that spirals can be seen in the scattered light emission of discs perturbed by inclined -- but not coplanar -- flybys. This implies, in agreement with other findings, that warped discs have spiral substructures that are visible in the scattered light. Scattered light observations in conjunction with kinematic observations of discs are therefore a powerful tool for definitively identifying discs that have undergone stellar flybys.

\section*{Acknowledgements}

VRP thanks the Science and Technology Facilities Council (STFC)
for a Ph.D. studentship. CL and CJC have been supported by STFC via the consolidated grant ST/W000997/1. CK acknowledges support from the European Union (ERC Starting Grant DiscEvol, project number 101039651) and from Fondazione Cariplo, grant No. 2022-1217. Views and opinions expressed are, however, those of the author(s) only and do not necessarily reflect those of the European Union or the European Research Council. Neither the European Union nor the granting authority can be held responsible for them.

The authors thank Antoine Alaguero and Marion Villenave for helpful discussions on scattered light, Andrés Izquierdo for support with tracing filament velocities in \textsc{discminer}, and Nicolás Cuello, Giuseppe Lodato and Andrew Winter for useful feedback. 

The simulations presented in this work were performed using the DiRAC Data Intensive service at Leicester (DiAL3), operated by the University of Leicester IT Services, which is part of the STFC DiRAC HPC Facility (\url{www.dirac.ac.uk}) within the RAC large project DISCSIM IV and Cambridge Service for Data Driven Discovery (CSD3). 

We acknowledge the use of \textsc{Splash} \citep{SPLASH_2007} for rendering and visualisation of figures. This work made use of the \texttt{python} packages \texttt{NumPy} \citep{harris2020array}, \texttt{SciPy} \citep{2020SciPy-NMeth}, \texttt{astropy} \citep{astropy:2013, astropy:2018, astropy:2022}, \texttt{sarracen} \citep{Sarracen} and \textit{Cmasher} \citep{cmasher}.

\section*{Data Availability}

Movies, a selection of dump files for each simulation and \textsc{mcfost} parameter files are available at \url{https://doi.org/10.5281/zenodo.22663318}. Full \textsc{Phantom} simulation output will be made available at reasonable request to the corresponding author.



\bibliographystyle{mnras}
\bibliography{references} 



\clearpage

\appendix

\section{Additional warp fitting output} \label{sec:warpfitter_extra}

In this section we present the full output of the fitting procedure performed by the \texttt{warpfitter} package for both CO and \textsuperscript{13}CO synthetic observations for the disc of each of the five flyby simulations \texttt{beta0}, \texttt{beta45}, \texttt{beta90}, \texttt{beta135} and \texttt{beta180} at 5402 yr after periastron. Full details on \texttt{warpfitter}'s least squares fitting procedure can be found in \citet{Winter_2025}. Note that the twist angle profiles $\gamma(R)$ shown here do not match those presented in \autoref{fig:gamma_warpfitter}, because in \autoref{fig:gamma_warpfitter} we utilised the \texttt{NumPy.unwrap} function to produce smooth and continuous profiles of $\gamma(R)$ that could be more easily compared with the profiles of $\gamma(R)$ extracted from the hydrodynamics.

\begin{figure*}
    \centering
    \includegraphics[width=\textwidth]{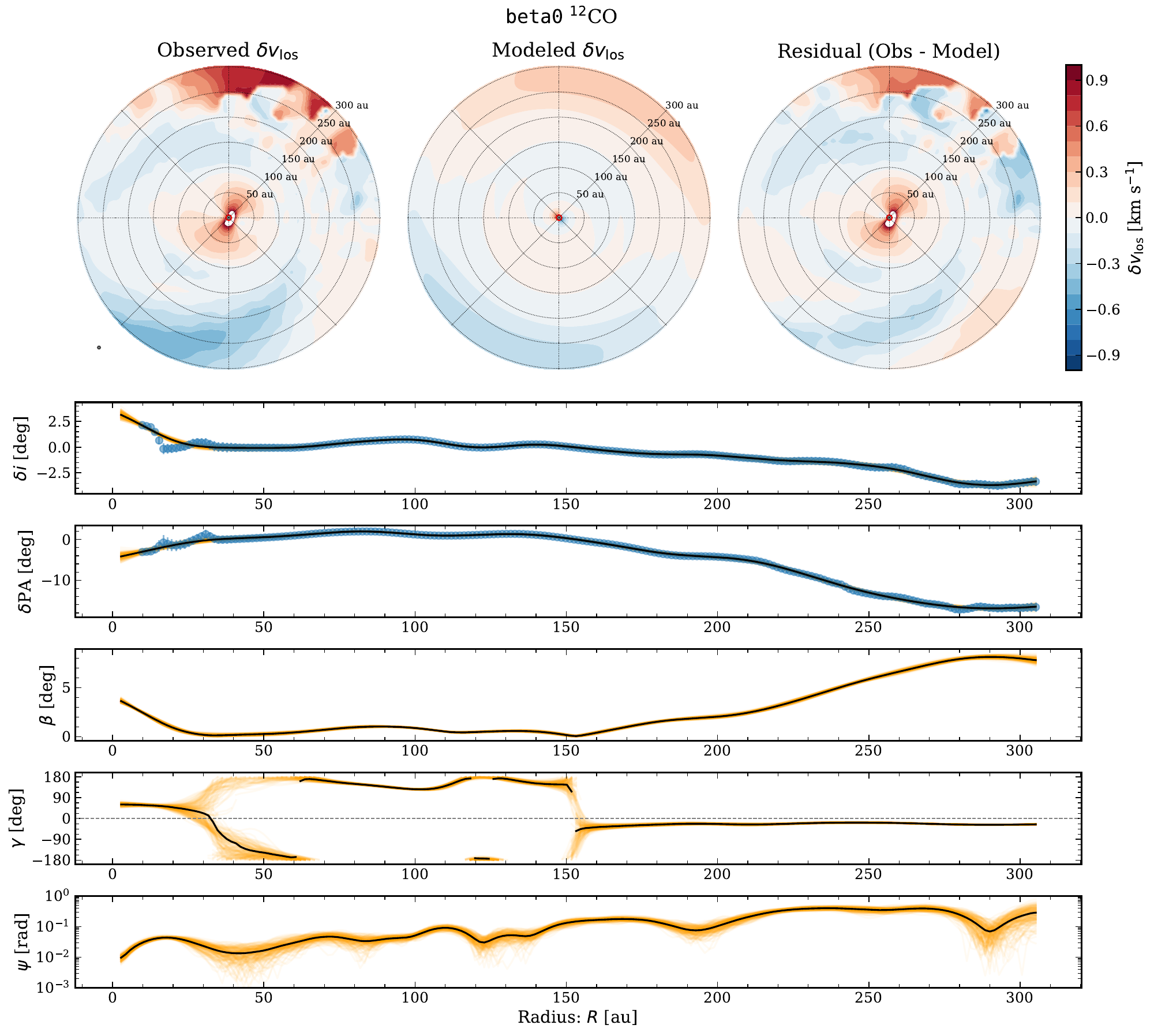}
    \caption{Top panels show (left to right): the LOS velocity residuals from the synthetic CO emission observation from the \texttt{beta0} simulation 5402 yr after periastron, as in \autoref{fig:CO_deproj_residuals};  the modelled $\delta v_\mathrm{los}$ field assumed by \texttt{warpfitter} by attributing all features in the residual velocity field to a warp; and the residuals when the modelled $\delta v_\mathrm{los}$ field is subtracted off the synthetic observation. The bottom panels show radial profiles of $\delta i$, $\delta\mathrm{PA}$, $\beta$, $\gamma$ and $\psi$ inferred from the \texttt{warpfitter} fitting procedure. Blue points and error bars in $\delta i$ and $\delta\mathrm{PA}$ come from \texttt{warpfitter}'s least squares fitting procedure. Faint orange lines show posterior distributions from \texttt{warpfitter}'s Gaussian process fit.}
    \label{fig:beta0_co_warpfitter}
\end{figure*}

\begin{figure*}
    \centering
    \includegraphics[width=\textwidth]{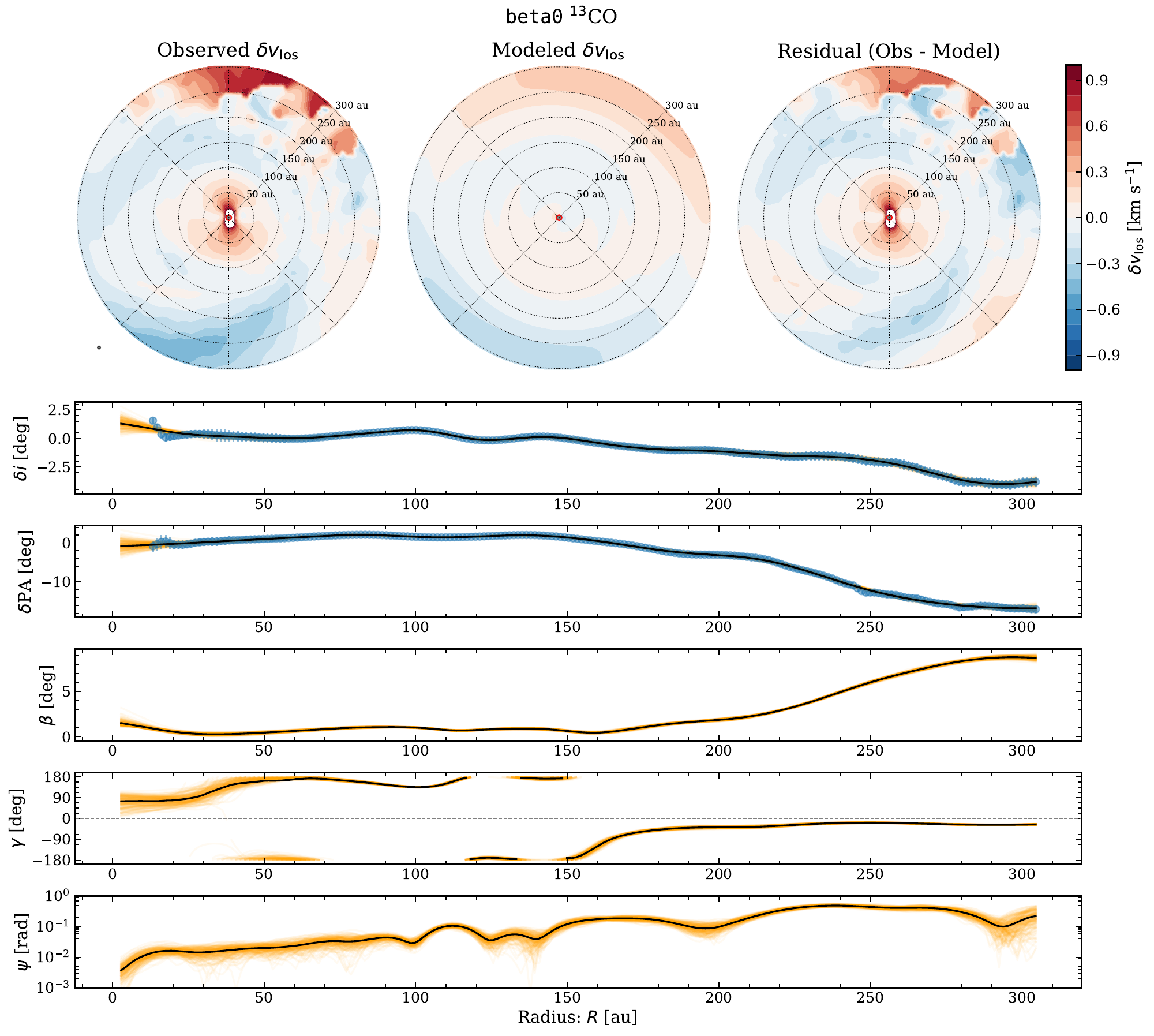}
    \caption{As in \autoref{fig:beta0_co_warpfitter}, but for \textsuperscript{13}CO.}
    \label{fig:beta0_13co_warpfitter}
\end{figure*}

\begin{figure*}
    \centering
    \includegraphics[width=\textwidth]{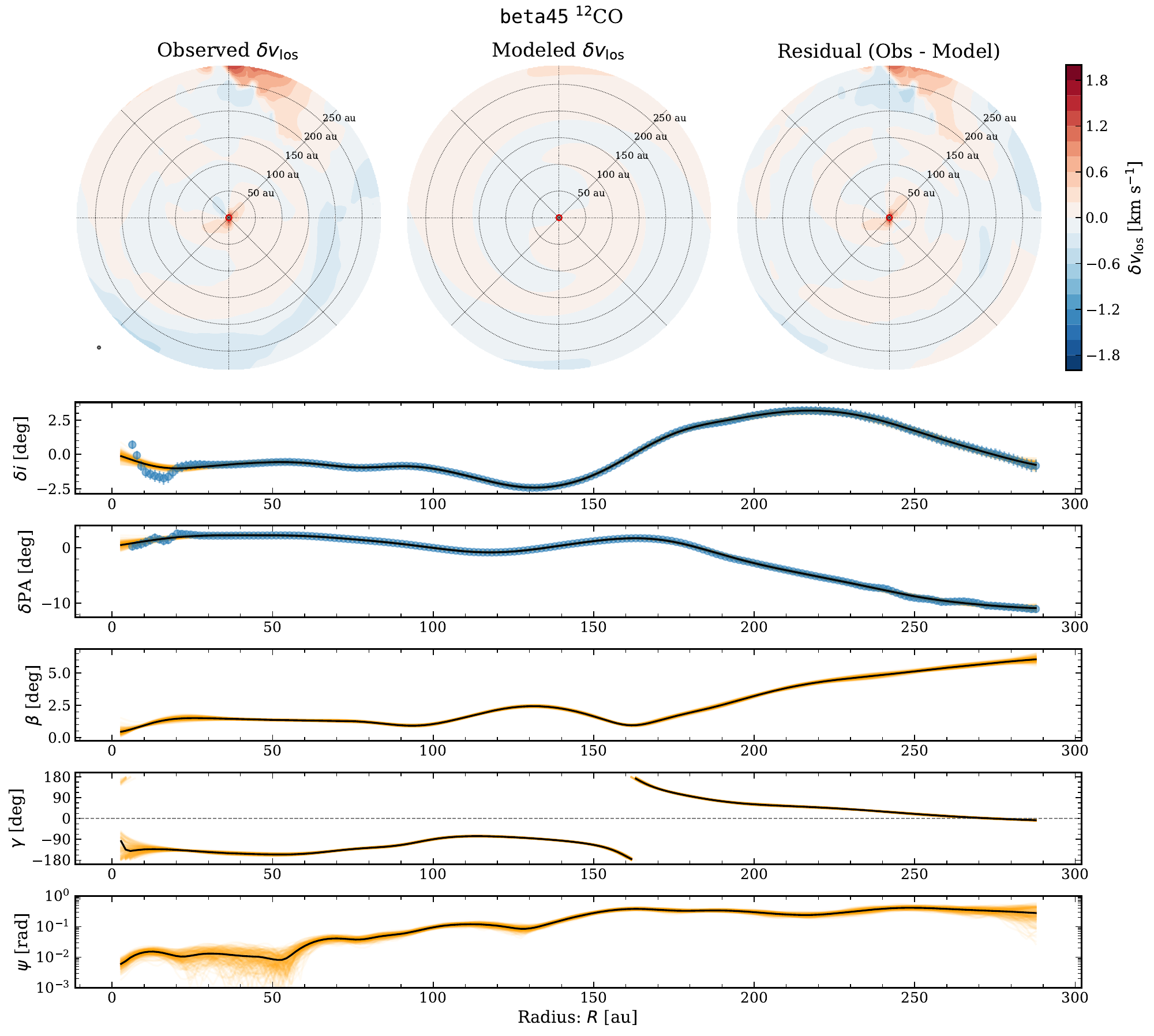}
    \caption{As in \autoref{fig:beta0_co_warpfitter}, but for the \texttt{beta45} simulation.}
    \label{fig:beta45_co_warpfitter}
\end{figure*}

\begin{figure*}
    \centering
    \includegraphics[width=\textwidth]{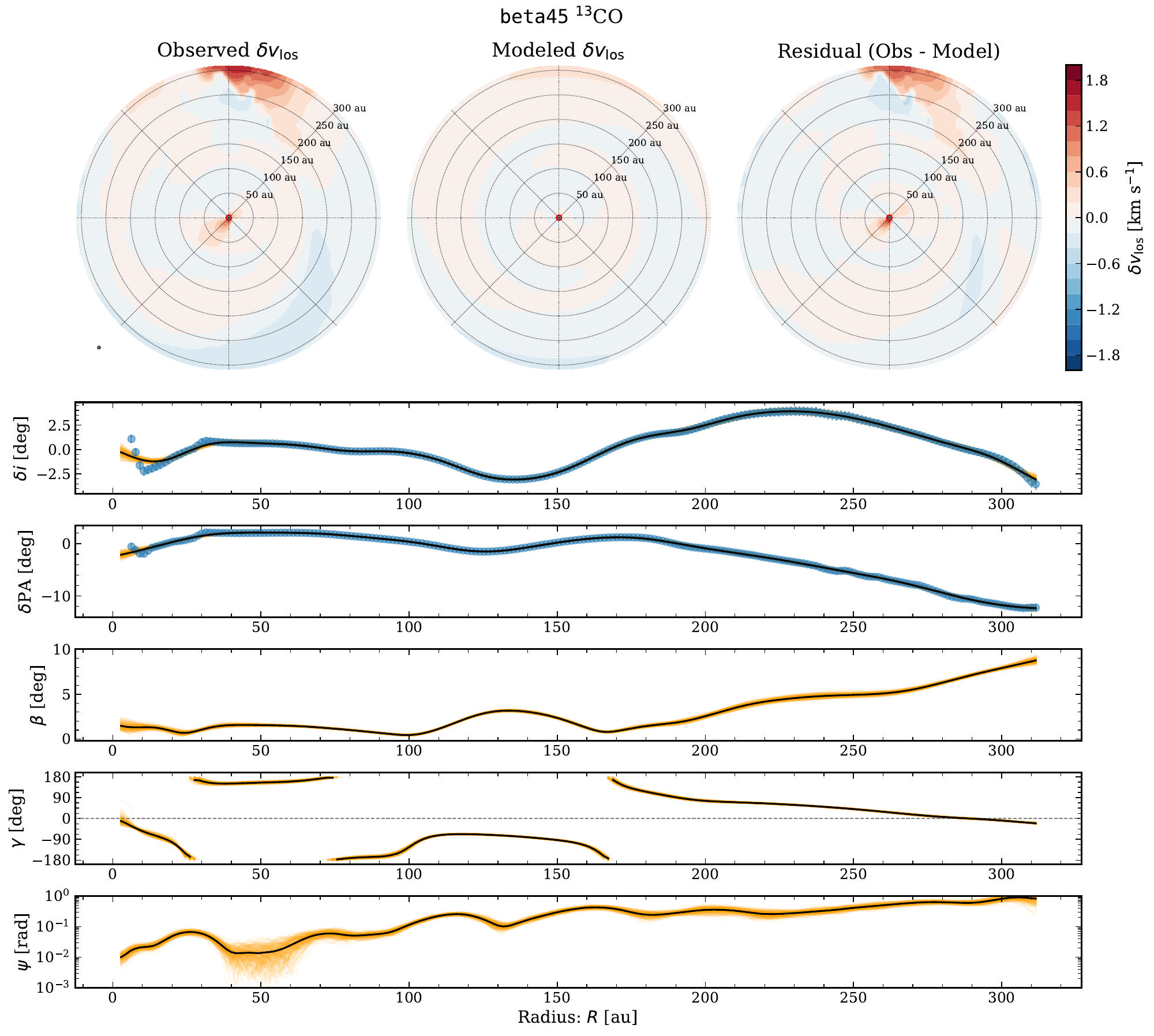}
    \caption{As in \autoref{fig:beta45_co_warpfitter}, but for \textsuperscript{13}CO.}
    \label{fig:beta45_13co_warpfitter}
\end{figure*}

\begin{figure*}
    \centering
    \includegraphics[width=\textwidth]{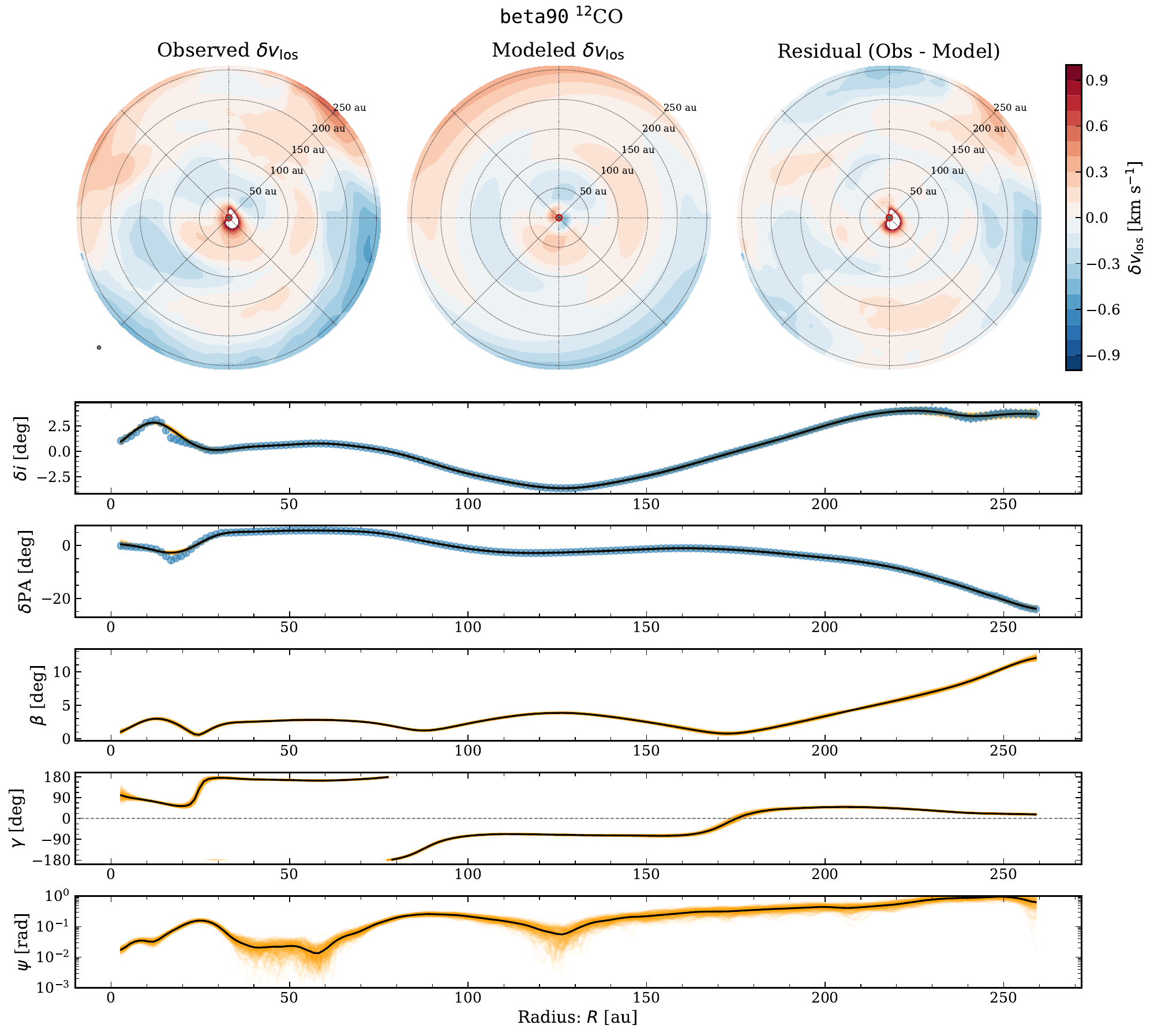}
    \caption{As in \autoref{fig:beta0_co_warpfitter}, but for the \texttt{beta90} simulation.}
    \label{fig:beta90_co_warpfitter}
\end{figure*}

\begin{figure*}
    \centering
    \includegraphics[width=\textwidth]{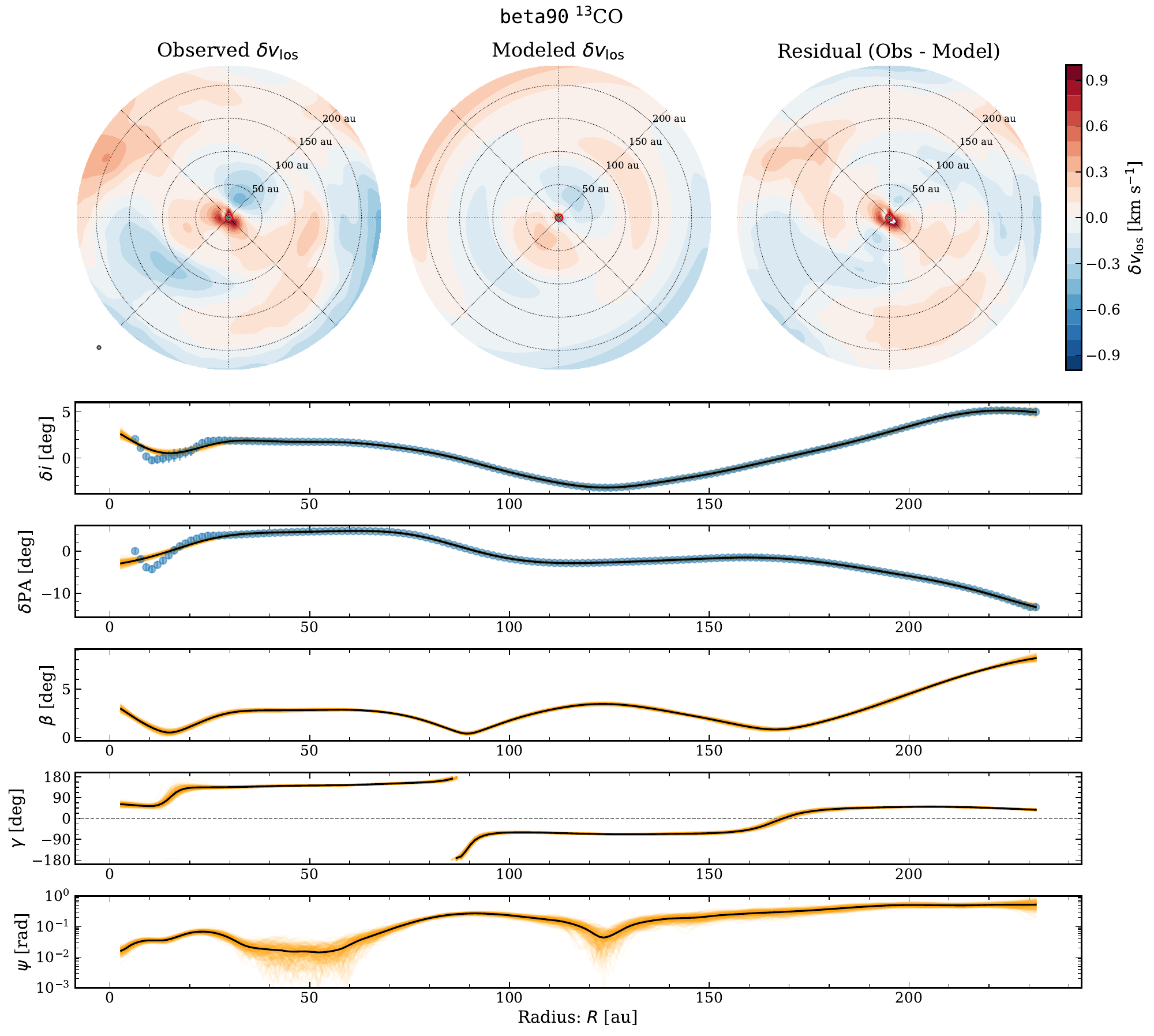}
    \caption{As in \autoref{fig:beta90_co_warpfitter}, but for \textsuperscript{13}CO.}
    \label{fig:beta90_13co_warpfitter}
\end{figure*}

\begin{figure*}
    \centering
    \includegraphics[width=\textwidth]{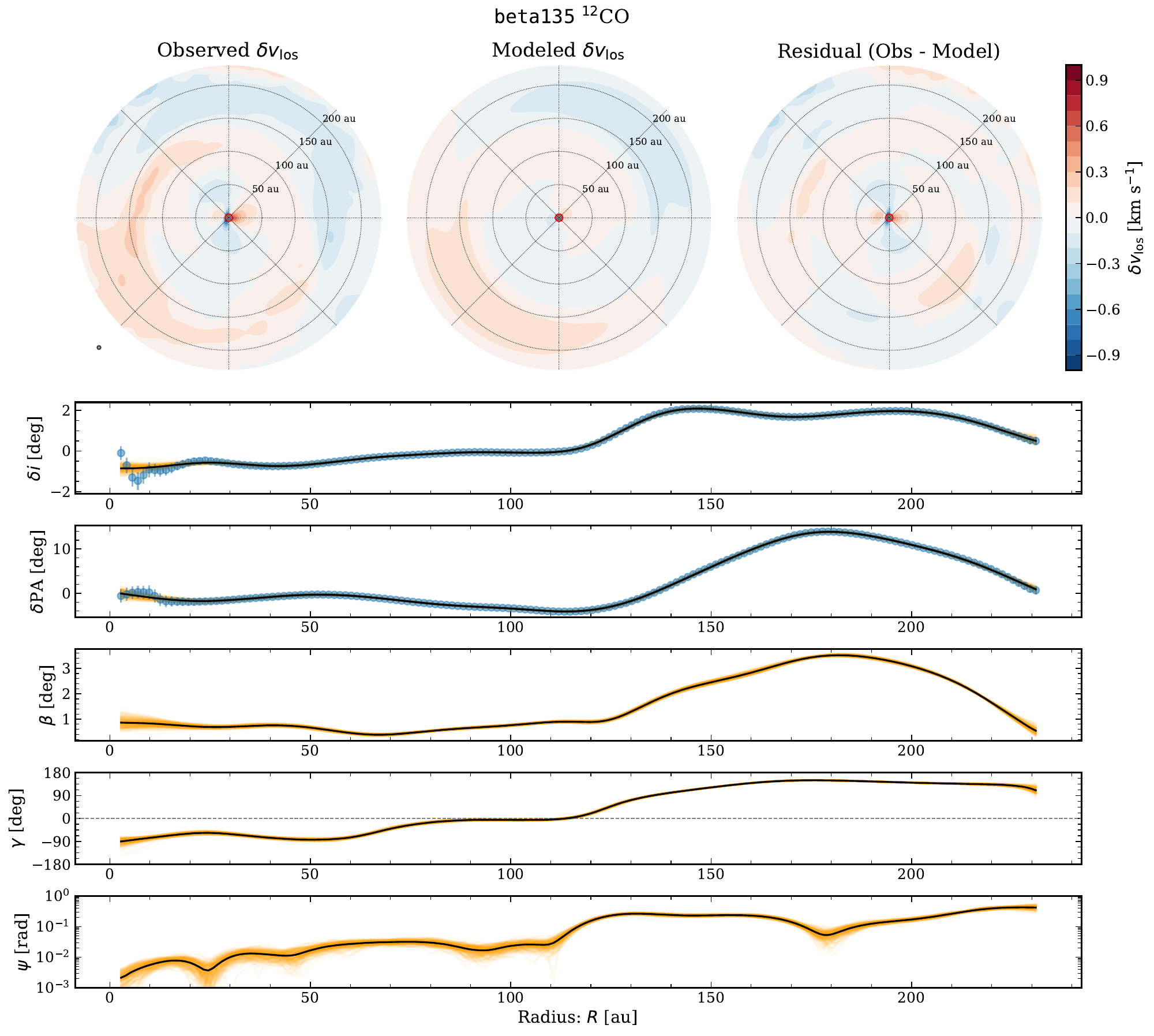}
    \caption{As in \autoref{fig:beta0_co_warpfitter}, but for the \texttt{beta135} simulation.}
    \label{fig:beta135_co_warpfitter}
\end{figure*}

\begin{figure*}
    \centering
    \includegraphics[width=\textwidth]{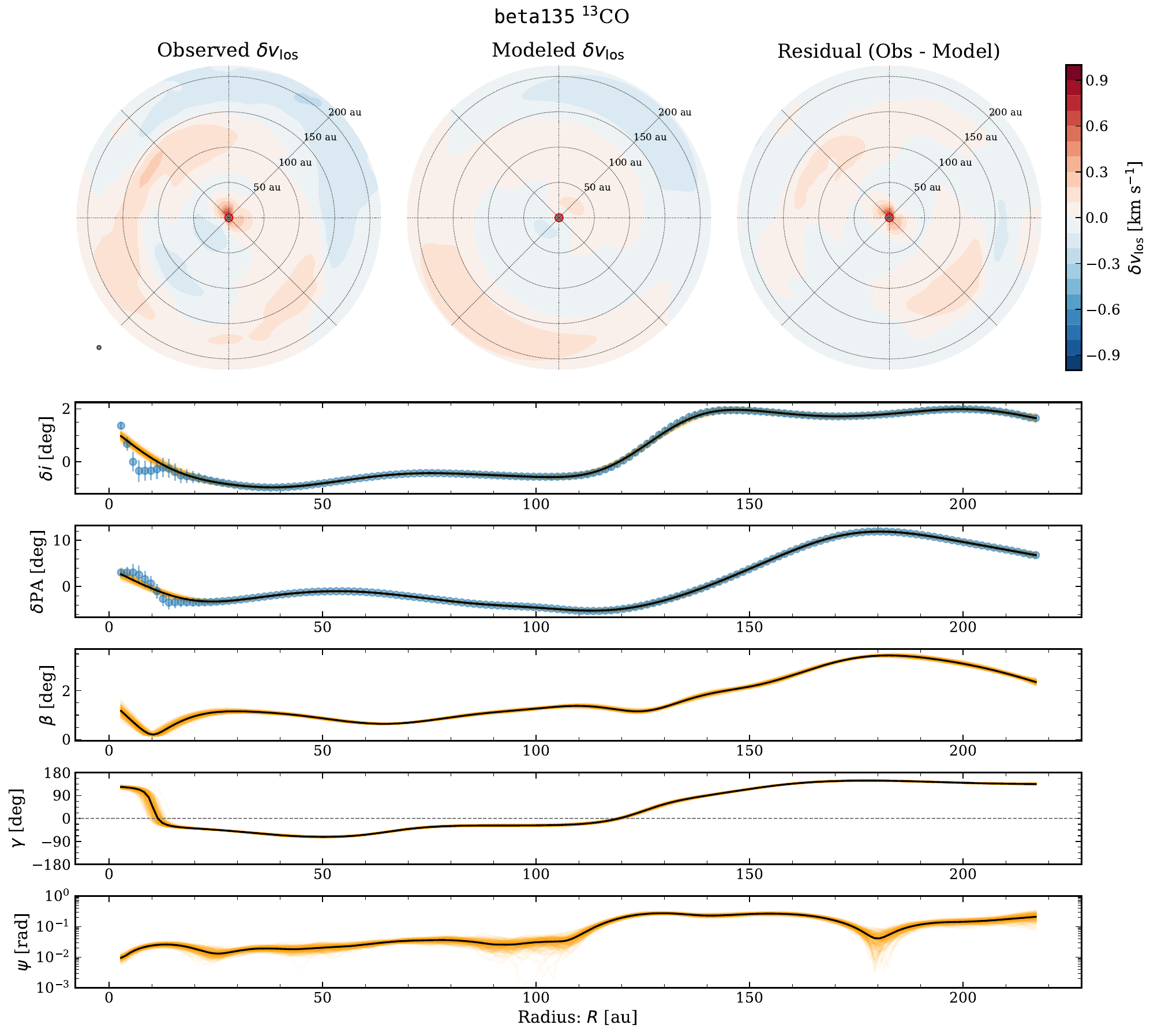}
    \caption{As in \autoref{fig:beta135_co_warpfitter}, but for \textsuperscript{13}CO.}
    \label{fig:beta135_13co_warpfitter}
\end{figure*}

\begin{figure*}
    \centering
    \includegraphics[width=\textwidth]{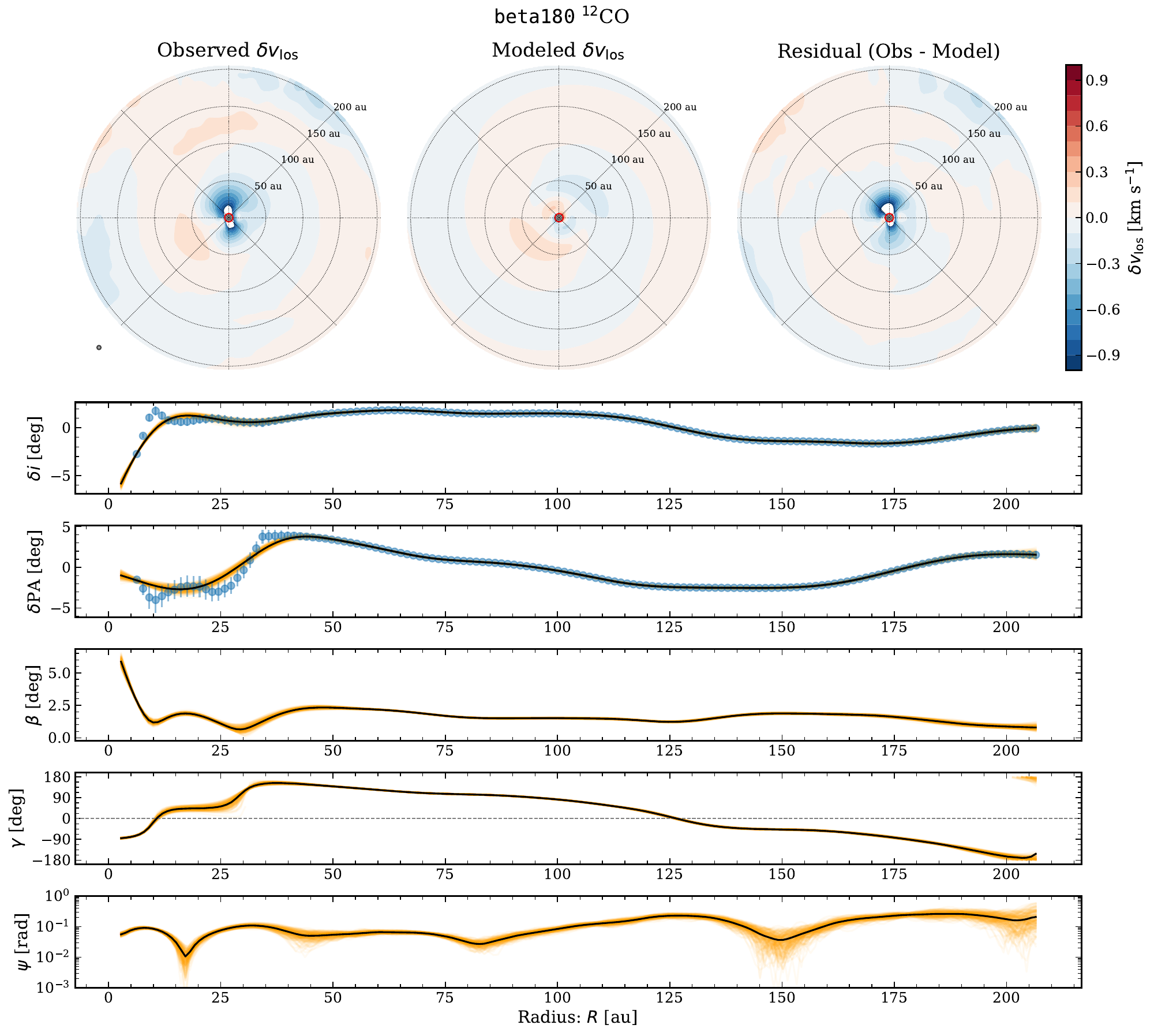}
    \caption{As in \autoref{fig:beta0_co_warpfitter}, but for the \texttt{beta180} simulation.}
    \label{fig:beta180_co_warpfitter}
\end{figure*}

\begin{figure*}
    \centering
    \includegraphics[width=\textwidth]{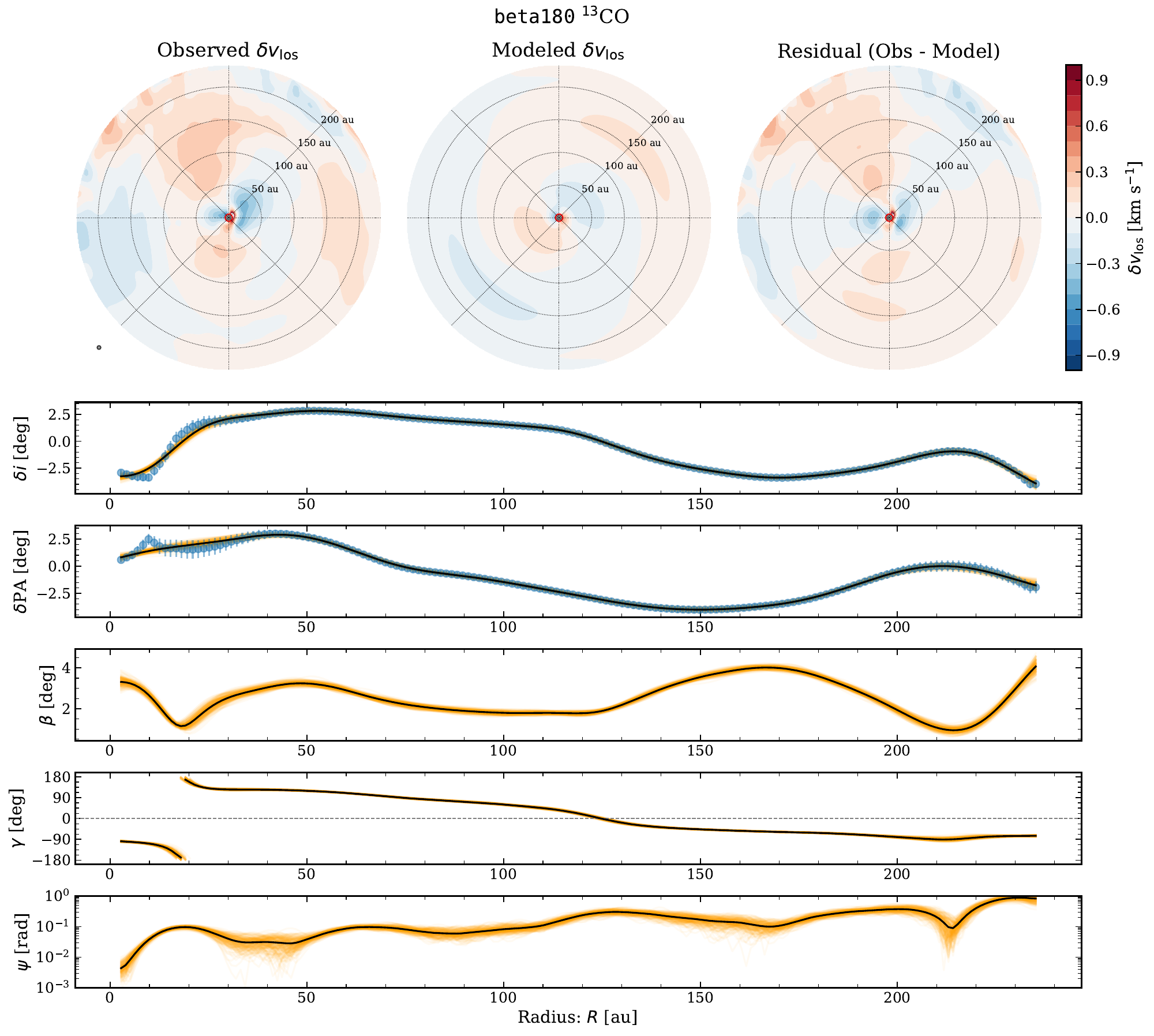}
    \caption{As in \autoref{fig:beta180_co_warpfitter}, but for \textsuperscript{13}CO.}
    \label{fig:beta180_13co_warpfitter}
\end{figure*}


\bsp	
\label{lastpage}
\end{document}

%% file: discminer_table_orientation.tex
\begin{table*}
\centering
\caption{Best-fit Keplerian stellar mass, systemic velocity and orientation parameters inferred with \textsc{discminer} for the CO and $^{13}$CO $J=3-2$ line emission of the five simulations at 5402 yr after periastron.}
\label{tab:discminer_table}
\begin{tabular}{lrrrrrrrrrrrrrr}
\hline
Simulation & \multicolumn{2}{c}{$M_\star$} & \multicolumn{2}{c}{$v_{\rm sys}$} & \multicolumn{2}{c}{$i$} & \multicolumn{2}{c}{PA} & \multicolumn{2}{c}{$x_c$} & \multicolumn{2}{c}{$y_c$} \\
 & CO & $^{13}$CO & CO & $^{13}$CO & CO & $^{13}$CO & CO & $^{13}$CO & CO & $^{13}$CO & CO & $^{13}$CO \\
 & \multicolumn{2}{c}{($M_\odot$)} & \multicolumn{2}{c}{(km/s)} & \multicolumn{2}{c}{($^\circ$)} & \multicolumn{2}{c}{($^\circ$)} & \multicolumn{2}{c}{(mas)} & \multicolumn{2}{c}{(mas)} \\
\hhline{===============}
\texttt{beta0} & 1.25 & 1.06 & 0.06 & 0.07 & 25.3 & 28.0 & 270.4 & 269.5 & 3.8 & -13.1 & 12.9 & 23.0 \\
\texttt{beta45} & 1.28 & 0.90 & 0.28 & 0.27 & 33.4 & 41.7 & 267.5 & 267.7 & 6.4 & 12.7 & -12.5 & -15.3 \\
\texttt{beta90} & 1.14 & 1.10 & 0.42 & 0.39 & 28.6 & 29.3 & 262.2 & 263.0 & -24.6 & -11.5 & 3.2 & -20.5 \\
\texttt{beta135} & 1.57 & 1.27 & 0.40 & 0.41 & 12.6 & 14.3 & 279.0 & 280.1 & -12.6 & -11.2 & -40.1 & -15.1 \\
\texttt{beta180} & 0.92 & 0.70 & 0.35 & 0.33 & 30.4 & 37.3 & 267.0 & 268.7 & 2.4 & 9.1 & -26.1 & 9.3 \\
\hline
\end{tabular}
\end{table*}